\documentclass[twocolumn,iop,numberedappendix,appendixfloats]{oja}

\usepackage[utf8]{inputenc}
\usepackage[british]{babel}

\usepackage{afterpage}
\usepackage{rotating}

\usepackage{amsmath}	% Advanced maths commands
 \usepackage{amssymb}	% Extra maths symbols
\usepackage{subcaption}  %subfigure
\usepackage{caption}

\usepackage{xcolor}
\definecolor{maroon}{cmyk}{0, 0.87, 0.68, 0.32}
\definecolor{halfgray}{gray}{0.55}
\definecolor{ipython_frame}{RGB}{207, 207, 207}
\definecolor{ipython_bg}{RGB}{247, 247, 247}
\definecolor{ipython_red}{RGB}{186, 33, 33}
\definecolor{ipython_green}{RGB}{0, 128, 0}
\definecolor{ipython_cyan}{RGB}{64, 128, 128}
\definecolor{ipython_purple}{RGB}{170, 34, 255}
\usepackage{float}
\usepackage{fontawesome}
\usepackage[colorlinks,allcolors=blue]{hyperref}
\usepackage{url}

\usepackage{listings}

\lstdefinelanguage{iPython}{
    morekeywords={access,and,break,class,continue,def,del,elif,else,except,exec,finally,for,from,global,if,import,in,is,lambda,not,or,pass,print,raise,return,try,while},%
    morekeywords=[2]{abs,all,any,basestring,bin,bool,bytearray,callable,chr,classmethod,cmp,compile,complex,delattr,dict,dir,divmod,enumerate,eval,execfile,file,filter,float,format,frozenset,getattr,globals,hasattr,hash,help,hex,id,input,int,isinstance,issubclass,iter,len,list,locals,long,map,max,memoryview,min,next,object,oct,open,ord,pow,property,range,raw_input,reduce,reload,repr,reversed,round,set,setattr,slice,sorted,staticmethod,str,sum,super,tuple,type,unichr,unicode,vars,xrange,zip,apply,buffer,coerce,intern},%
    sensitive=true,%
    morecomment=[l]\#,%
    morestring=[b]',%
    morestring=[b]",%
    morestring=[s]{'''}{'''},% used for documentation text (mulitiline strings)
    morestring=[s]{"""}{"""},% added by Philipp Matthias Hahn
    morestring=[s]{r'}{'},% `raw' strings
    morestring=[s]{r"}{"},%
    morestring=[s]{r'''}{'''},%
    morestring=[s]{r"""}{"""},%
    morestring=[s]{u'}{'},% unicode strings
    morestring=[s]{u"}{"},%
    morestring=[s]{u'''}{'''},%
    morestring=[s]{u"""}{"""},%
    literate=
    {á}{{\'a}}1 {é}{{\'e}}1 {í}{{\'i}}1 {ó}{{\'o}}1 {ú}{{\'u}}1
    {Á}{{\'A}}1 {É}{{\'E}}1 {Í}{{\'I}}1 {Ó}{{\'O}}1 {Ú}{{\'U}}1
    {à}{{\`a}}1 {è}{{\`e}}1 {ì}{{\`i}}1 {ò}{{\`o}}1 {ù}{{\`u}}1
    {À}{{\`A}}1 {È}{{\'E}}1 {Ì}{{\`I}}1 {Ò}{{\`O}}1 {Ù}{{\`U}}1
    {ä}{{\"a}}1 {ë}{{\"e}}1 {ï}{{\"i}}1 {ö}{{\"o}}1 {ü}{{\"u}}1
    {Ä}{{\"A}}1 {Ë}{{\"E}}1 {Ï}{{\"I}}1 {Ö}{{\"O}}1 {Ü}{{\"U}}1
    {â}{{\^a}}1 {ê}{{\^e}}1 {î}{{\^i}}1 {ô}{{\^o}}1 {û}{{\^u}}1
    {Â}{{\^A}}1 {Ê}{{\^E}}1 {Î}{{\^I}}1 {Ô}{{\^O}}1 {Û}{{\^U}}1
    {œ}{{\oe}}1 {Œ}{{\OE}}1 {æ}{{\ae}}1 {Æ}{{\AE}}1 {ß}{{\ss}}1
    {ç}{{\c c}}1 {Ç}{{\c C}}1 {ø}{{\o}}1 {å}{{\r a}}1 {Å}{{\r A}}1
    {€}{{\EUR}}1 {£}{{\pounds}}1
    {^}{{{\color{ipython_purple}\^{}}}}1
    {=}{{{\color{ipython_purple}=}}}1
    {+}{{{\color{ipython_purple}+}}}1
    {-}{{{\color{ipython_purple}-}}}1
    {*}{{{\color{ipython_purple}$^\ast$}}}1
    {/}{{{\color{ipython_purple}/}}}1
    {+=}{{{+=}}}1
    {-=}{{{-=}}}1
    {*=}{{{$^\ast$=}}}1
    {/=}{{{/=}}}1,
    literate=
    *{-}{{{\color{ipython_purple}-}}}1
     {?}{{{\color{ipython_purple}?}}}1,
    identifierstyle=\color{black}\ttfamily,
    commentstyle=\color{ipython_cyan}\ttfamily,
    stringstyle=\color{ipython_red}\ttfamily,
    keepspaces=true,
    showspaces=false,
    showstringspaces=false,
    rulecolor=\color{ipython_frame},
    frameround={t}{t}{t}{t},
    numbers=none,
    numberstyle=\tiny\color{halfgray},
    backgroundcolor=\color{ipython_bg},
    basicstyle=\ttfamily\footnotesize,
    columns=fullflexible,
    keywordstyle=\color{ipython_green}\ttfamily,
}

\usepackage{array,multirow,graphicx}

\newcommand{\be}{\begin{equation}}
\newcommand{\ee}{\end{equation}}
\newcommand{\bea}{\begin{eqnarray}}
\newcommand{\eea}{\end{eqnarray}}

\newcommand{\react}{\textsc{REACT}}
\newcommand{\cosmopower}{\textsc{COSMOPOWER}}
\newcommand{\reactemufr}{\textsc{REACTEMU-FR}}
\newcommand{\bcemu}{\textsc{BCEMU}}
\newcommand{\halofit}{\textsc{HALOFIT}}
\newcommand{\HMcode}{\textsc{HMCODE}}

\newcommand{\orcid}[1]{\href{https://orcid.org/#1}{\includegraphics[width=10pt]{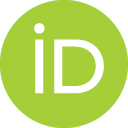}}}

\begin{document}

\journalinfo{The Open Journal of Astrophysics}
\submitted{submitted XXX; accepted YYY}

\title{On the degeneracies between baryons, massive neutrinos and $f(R)$ gravity \\ in Stage IV cosmic shear analyses\vspace{-1.5cm}}

\shorttitle{Baryons, neutrinos and $f(R)$ gravity from Stage IV cosmic shear}
\shortauthors{A. Spurio Mancini \& B. Bose}

% The list of authors, and the short list which is used in the headers.
\author{A. Spurio Mancini$^{1}$ \orcid{0000-0001-5698-0990} \thanks{E-mail: \href{a.spuriomancini@ucl.ac.uk}{a.spuriomancini@ucl.ac.uk}}}
\author{B. Bose$^{2,3,4}$ \orcid{0000-0003-1965-8614} \vspace{1em}}

\affiliation{$^1$ Mullard Space Science Laboratory, University College London, Holmbury St. Mary, Dorking, Surrey, RH5 6NT, UK}
\affiliation{$^{2} $Institute for Astronomy, University of Edinburgh, Royal Observatory, Blackford Hill, Edinburgh, EH9 3HJ, UK} 
\affiliation{$^{3}$Basic Research Community for Physics e.V., Mariannenstraße 89, Leipzig, Germany}
\affiliation{$^{4}$D\'epartement de Physique Th\'eorique, Universit\'e de Gen\`eve, 24 quai Ernest Ansermet, 1211 Gen\`eve 4, Switzerland}

\vspace{2.5cm}

\begin{abstract}
Modelling nonlinear structure formation is essential for current and forthcoming cosmic shear experiments. We combine the halo model reaction formalism, implemented in the \react{} code, with the \cosmopower{} machine learning emulation platform, to develop and publicly release \reactemufr{} \href{https://github.com/cosmopower-organization/reactemu-fr}{\faicon{github}}, a fast and accurate nonlinear matter power spectrum emulator for $f(R)$ gravity with massive neutrinos. Coupled with the state-of-the-art baryon feedback emulator \bcemu{}, we use \reactemufr{} to produce Markov Chain Monte Carlo forecasts for a cosmic shear experiment with typical Stage IV specifications. We find that the inclusion of highly nonlinear scales (multipoles between $1500\leq \ell \leq 5000$) only mildly improves constraints on most standard cosmological parameters (less than a factor of 2). In particular, the necessary modelling of baryonic physics effectively damps most constraining power on the sum of the neutrino masses and modified gravity at $\ell \gtrsim 1500$. Using an approximate baryonic physics model produces mildly improved constraints on cosmological parameters which remain unbiased at the $1\sigma$-level, but significantly biases constraints on baryonic parameters at the $> 2\sigma$-level. 
\vspace{1em}
\end{abstract}


\maketitle

% ================================================
%                   INTRODUCTION
% ================================================
\section{Introduction}
\label{sec:introduction}
% Constraining gravity with Stage IV WL
One of the main goals of forthcoming Stage IV cosmic shear surveys, such as \textit{Euclid} \citep{EUCLID:2011zbd}\footnote{https://www.euclid-ec.org/}, the Vera \textit{Rubin} Observatory's Legacy Survey of Space and Time \citep[VRO-LSST,][]{LSST:2008ijt}\footnote{https://www.lsst.org/} and the Nancy Grace \textit{Roman} Telescope \citep{Roman}\footnote{https://roman.gsfc.nasa.gov/}, is to perform precise and accurate tests of gravity on cosmological scales. Constraints on deviations from General Relativity, the gravity theory underlying our standard LCDM model of cosmology, will particularly benefit from measurements of the cosmic shear power spectrum from unprecedentedly large numbers of galaxies. Contributions to the cosmic shear power spectrum from small, nonlinear scales, will be of the utmost importance to distinguish between competing gravity models with different nonlinear screening mechanisms \citep[see \textit{e.g.}][]{Koyama_2016, Harnois-Deraps:2022bie}.

% Theoretical requirements
On the theoretical end, the achievement of this goal hinges on various factors. We list some of the primary criteria below:
\begin{enumerate}
    \item we require accurate prescriptions for the key observables of Stage IV surveys, at both linear and nonlinear scales; 
    \item we require these prescriptions to be computationally efficient, to use them within rigorous Bayesian analysis pipelines;
    \item we need to theoretically model all relevant and known physical phenomena in order to distinguish any potential new physics.    
\end{enumerate}

% Available prescriptions
So far, these criteria have only been met for a subset of theoretical descriptions of gravity and cosmology, and all suffer from various limitations. Emulators trained on $N$-body simulations \citep[\textit{e.g.}][]{Lawrence:2017ost,Euclid:2018mlb,Euclid:2020rfv,Angulo:2020vky,Ramachandra:2020lue,Arnold:2021xtm} are restricted to the parameter space probed by their training simulations, and cannot be safely extended without running new simulations. Analytic prescriptions \citep[\textit{e.g.}][]{Takahashi:2012em,Zhao:2013dza,Mead:2016zqy,Winther:2019mus,Mead:2020vgs} are also restricted to the simulations which they have originally been fit to.

% ReACT
A method potentially capable of overcoming some of these issues is the halo model reaction, a theoretically flexible framework introduced in \cite{Cataneo:2018cic} \citep[see also][for initial motivations]{Mead:2016ybv} based on 1-loop perturbation theory and the halo model \citep[see \textit{e.g.}][for a recent review]{Asgari_23}, shown to be consistent with $N$-body simulations at the $\sim2\%$ level down to $k\sim3h/{\rm Mpc}$ for a wide range of non-standard cosmological scenarios \citep{Srinivasan:2021gib,Bose:2021mkz,Carrilho:2021rqo,Parimbelli:2022pmr,Euclid:2022qde}. However, this method is not as accurate as $N$-body based emulators \citep[see \textit{e.g.}][]{Arnold:2021xtm}, nor is the associated code \react{}~\citep[\href{https://github.com/nebblu/ACTio-ReACTio}{\faicon{github}}]{Bose:2020wch,Bose:2022vwi} as fast as current emulators. 

% CosmoPower
This being said, the sacrifice in accuracy of the halo model reaction may be a non-issue in real data analyses, once we take into account relevant physical effects such as baryon feedback processes and massive neutrinos. Furthermore, the computational inefficiency of \react{} can be circumvented by employing the recent \cosmopower{} machine learning platform \citep[\href{https://github.com/alessiospuriomancini/cosmopower}{\faicon{github}}]{SpurioMancini:2021ppk}, which trains fast neural network emulators of key cosmological quantities starting from a set of `slow' theoretical predictions. Once trained, these emulators can be used in Bayesian inference pipelines to replace the call to the original, computationally intensive prescription, massively accelerating parameter estimation. Developing similar accelerated pipelines is critical to the success of forthcoming Stage IV cosmic shear analyses, as these will have to be carried out over unprecedentedly large parameter spaces for accurate modelling of baryons, massive neutrinos, systematics and deviations from LCDM. 

% In this paper
In this paper we present \reactemufr{}, an emulator of the nonlinear effects on the matter power spectrum caused by massive neutrinos in Hu-Sawicki $f(R)$ gravity \citep{Hu:2007nk}, a popular non-standard model of gravity being considered by Stage IV surveys \citep[\textit{e.g.}][]{Amendola:2016saw,Ramachandra:2020lue}. This emulator is constructed by \cosmopower{} using \react{} predictions, demonstrating the ease by which non-standard model emulators can be generated using the two codes, and seamlessly integrated into analysis pipelines. We use this emulator to forecast the constraining power on cosmology and gravity, as well as to investigate parameter degeneracies using cosmic shear, in the context of Stage IV surveys. We check the capability of such surveys to detect new physics given a full host of secondary physical phenomena including baryon effects and massive neutrinos. Conversely, we also investigate the importance of including all non-standard physical effects in order to remain unbiased in the final parameter constraints. 

% Structure of the paper
This paper is outlined as follows. In \autoref{sec:theory} we present and validate \reactemufr{} and discuss baryonic effects and cosmic shear. In \autoref{sec:forecasts} we perform forecasts for a Stage IV survey using different sets of degrees of freedom. We summarise our findings and conclude in \autoref{sec:conclusions}.

% ================================================
%                   THEORY
% ================================================
\section{Modelling cosmic shear with emulators}\label{sec:theory}

\subsection{Massive neutrinos and $f(R)$ gravity: \reactemufr{}}
% Massive neutrinos
Massive neutrinos entered the regime of standard physics two decades ago with the measurements of flavour oscillations \citep{Super-Kamiokande:1998qwk,SNO:2003bmh}, and were shown to have a significant effect on cosmological structure formation \citep{Lesgourgues:2006nd}. They are usually parameterised in terms of the combined neutrino mass, $\sum m_\nu$. Neutrino oscillation experiments have placed a lower bound on this parameter of $\sum m_\nu \geq 0.06~{\rm eV}$. There is also an upper bound from the KATRIN collaboration of $\sum m_\nu \leq 0.8~{\rm eV}$ \citep{KATRIN:2021uub}.

% f(R)
Modified gravity theories are extensions of General Relativity developed to deviate from standard gravity on cosmological scales, where it remains largely untested. These models can provide interesting solutions to fundamental physical problems \citep[see \textit{e.g.}][for a recent review]{Bernardo:2022cck}. In this paper we select a specific theory of modified gravity - the popular Hu-Sawicki $f(R)$ model \citep{Hu:2007nk}. In the $f(R)$ class of models, the Einstein-Hilbert action is generalised to include an arbitrary  function of the scalar curvature, $R$. The Hu-Sawicki form is given by
\begin{equation}
f(R)  = -m^2 \frac{c_1 (R/m^2)^n}{c_2(R/m^2)^n+1} \, , 
\label{husawicki}
\end{equation}
where we choose $n=1$ and $m^2=\Omega_{\rm m} H_0^2$. $c_1$ and $c_2$ are parameters of the theory, and as is commonly done, we reparametrise these in terms of the the value of $f_{\rm R} = d f(R)/dR$ today
\begin{equation}
f_{\rm R_0} = - \frac{c_1}{c_2^2} \left( \frac{m^2}{\bar{R}_0} \right)^2 \, ,
\end{equation}
$\bar{R}_0$ being the background scalar curvature today. In the following we take the effective LCDM-limit  to be $|f_{\rm R_0}|=10^{-9}$. We have checked that this value induces effects in the observable of interest much smaller than all other sources of error, making it indistinguishable from the true LCDM limit of $f_{\rm R_0} = 0$.

This model predicts non-trivial scale dependencies in the growth of structure. It has been shown to have degeneracies with massive neutrinos \citep{Mead:2016zqy,Wright:2019qhf}. The degree of deviation from LCDM in this model is parameterised by the value of the additional scalar degree of freedom today, $f_{\rm R_0}$, with the LCDM limit of the theory given by $f_{\rm R_0}\rightarrow 0$. Current cosmological bounds on $f_{\rm R_0}$  are of $\mathcal{O}(10^{-5})$ \citep{Cataneo:2014kaa,Lombriser:2014dua,Desmond:2020gzn,Brax:2021wcv}. 

% emulated quantities
We train an emulator for the $f(R)+\nu$-boost, \textit{i.e.} the nonlinear modification to a LCDM spectrum caused by $f(R)$ and massive neutrinos, as a function of Fourier mode $k$ and scale factor $a$: 
\begin{align}
    B^{f(R)+\nu}(k,a) \equiv \frac{P_{\rm NL}^{f(R)+\nu}(k,a)}{P_{\rm NL}^{\rm \Lambda CDM} (k,a)} \, ,
    \label{eq:boostfR}
\end{align}
where $a$ is the scale factor and $P_{\rm NL}^{f(R)+\nu}$ and $P_{\rm NL}^{\rm \Lambda CDM}$ are the nonlinear matter power spectra in the $f(R)$ with massive neutrino cosmology and LCDM model, respectively. The nonlinear power spectrum under the halo model reaction prescription \citep{Cataneo:2018cic} is given by
\begin{align}
    P^{f(R)+\nu}_{\rm NL} (k, a) = \mathcal{R}(k,a) \times P^{\rm pseudo}_{\rm NL} (k, a) \, .
    \label{eq:nlpshmr}
\end{align}
$\mathcal{R}$ is the halo model reaction and $P^{\rm pseudo}_{\rm NL}$ is the pseudo power spectrum. The latter is defined as originating from a LCDM cosmology whose initial conditions are tuned such that the linear clustering matches the non-standard cosmology at a target redshift. In practice, we can construct this by supplying prescriptions such as \halofit{} \citep{Takahashi:2012em} or \HMcode{} \citep{Mead:2020vgs} with a modified linear power spectrum \citep[see][for details]{Bose:2020wch}.

% reaction
$\mathcal{R}$ encodes nonlinear corrections to the pseudo power spectrum due to $f(R)$ and massive neutrinos -- in essence, it is a ratio between halo model predictions for the non-standard cosmology and a pseudo cosmology, with a correction factor coming from a calibration with 1-loop perturbation theory at quasi-nonlinear scales. We refer the reader to \cite{Bose:2021mkz} for more details, and we follow this reference in our modelling of $\mathcal{R}$. Note that the background cosmology is assumed to be LCDM, with the effects of modified gravity and massive neutrinos only entering at the level of the perturbations.  

% emulator specifics
In this paper we use \HMcode{} \citep{Mead:2020vgs} to model the pseudo and LCDM power spectrum appearing in \autoref{eq:boostfR} and \autoref{eq:nlpshmr}. To compute the halo model reaction we use the \react{} code \citep{Bose:2022vwi}. Our \cosmopower{} emulator, \reactemufr{}, is then built on \autoref{eq:boostfR} computed within the parameter ranges in \autoref{tab:all_params}, sampled with a Latin Hypercube implemented using the \textsc{pyDOE} library \href{https://github.com/tisimst/pyDOE/}{\faicon{github}}, which allows for essentially instantaneous creations of Latin Hypercube grids. The ranges are chosen as the rough $5\sigma$ bounds around the Planck 2018 \citep{Aghanim:2018eyx} best fit values including a varying $\sum m_\nu$. We take an upper bound on $f_{\rm R_0}$ to be the $3\sigma$ limit found in the cosmic shear forecast of \cite{Bose:2020wch} using only linear scales. These represent very conservative ranges, with any deviation outside these being in strong disagreement with current observations. Note that the emulator can always be trained on extended ranges as the halo model reaction effectively has no prior range. The authors of \cite{Bose:2020wch} found the pseudo prescription of the earlier HMCODE \citep{Mead:2016zqy} to be insufficiently accurate for highly nonlinear scales in Stage-IV contexts. Here we use the more accurate version of \cite{Mead:2020vgs}, and we model the boost, which is far more accurate as the ratio of \HMcode{} predictions reduces systematics significantly \citep{Cataneo:2018cic}. Given that we wish to primarily investigate the bias incurred from incomplete baryonic physics modelling, we believe this prescription is sufficiently accurate for our purposes. \reactemufr{} is deemed to have an accuracy of $\sim 3\%$ at scales of $k\leq 1h/{\rm Mpc}$ within its prior range as shown by comparisons with emulators and $N$-body simulations \citep{Cataneo:2018cic,Bose:2021mkz,Arnold:2021xtm}, with worsening accuracy for larger deviations from LCDM. For reference, for $|f_{\rm R_0}| = 10^{-6}$ and $\sum m_\nu \leq 0.1$, the boost was shown to be $2\%$ consistent with $N$-body measurements \citep{Bose:2021mkz,Arnold:2021xtm}.

\begin{figure}
    \centering
    \includegraphics[width=\columnwidth]{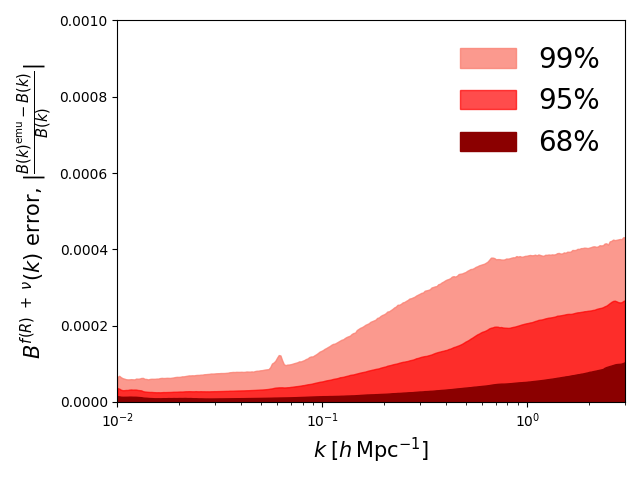}
    \caption{Quantiles of the relative accuracy of \reactemufr{} on $\sim 2 \cdot 10^4$ testing samples.}
    \label{fig:accuracy}
\end{figure}

% testing CosmoPower emulator
We train a \cosmopower{} emulator on $\sim10^5$ \react{} predictions and test its performance on $\sim 2 \cdot 10^4$ testing samples, finding $\leq 0.05\%$ residuals across the entire emulated $k$-range $k \in [0.01, 3]h$/Mpc, \textit{i.e.} the entire validity range of \react{}. \autoref{fig:accuracy} shows a quantile plot of the accuracy of our emulator on the testing set. This accuracy level is more than an order of magnitude bigger than that used in \citet{SpurioMancini:2021ppk} to validate their matter power spectrum emulators, which were shown to be comfortably accurate to guarantee unbiased emulated Stage IV posterior contours. The speed up factor provided by this emulation is above $\mathcal{O}(10^4)$ as the standard boost computation requires multiple Boltzmann solver calls, multiple \HMcode{} calls and a call to \react{}, with a typical boost taking up to $\sim 25~{\rm s}$ to compute compared to \reactemufr{}'s $\sim 1~{\rm ms}$. 

Note that the procedure of generating the training spectra is a highly parallelisable one, since every sample is completely independent from any other.  In our case, we used 400 cores to generate our $\sim 10^5$ training samples, which took approximately 10 hours in total, while the training of the emulator took approximately 1 hour on an NVIDIA T4 Graphics Processing Unit, as available for free on \href{http://colab.research.google.com/}{Google Colab}.

\begin{table*}
  \renewcommand{\arraystretch}{2}
  \setlength{\tabcolsep}{3.5pt}
\centering
\caption{Varied cosmological, beyond-LCDM and baryonic parameters at $z=0$, with prior ranges and fiducial values. The cosmological parameter fiducials are taken to be the Planck 2018 best-fits, while the baryonic parameter fiducials are the default values of \bcemu{}.}
\begin{tabular}{| l| l | l | l | l | l | l | l | l | l | l | l | l | l | l | }
\hline  
& $\Omega_{\rm m}$  & $\Omega_{\rm b}$  & $h$ & $n_{\rm s}$ & $A_{\rm s}  10^9 $  & $\sum m_\nu$ & $|f_{\rm R_0}|$ & $\log_{10} M_{\rm c}$ & $\mu$ & $\theta_{\rm ej}$  & $\gamma$  & $\delta$ & $\eta$ & $\eta_{\delta}$  \\ \hline 
{\bf Lower}  & 0.2899 & 0.04044 & 0.638 & 0.9432 & 1.9511 & 0 & $10^{-10}$        & 11.4  & 0.2  & 2.6   & 1.3 &  3.8 & 0.085 & 0.085 \\
{\bf Upper}  & 0.3392 & 0.05686 & 0.731 & 0.9862 & 2.2669 & 0.1576 & $10^{-5.46}$ & 14.6  & 1.8  & 7.4   & 3.7 &  10.2 & 0.365 & 0.365 \\ 
{\bf Fiducial} & 0.3164 & 0.04966 & 0.671 & 0.9647 & 2.1031 & 0.06 & See main text &13.32 & 0.93 & 4.235 & 2.25&  6.4 & 0.15 & 0.14 \\ 
\hline
\end{tabular}
\label{tab:all_params}
\end{table*}
\begin{table}
  \renewcommand{\arraystretch}{2}
  \setlength{\tabcolsep}{3.5pt}
\centering
\caption{Priors and fiducial values for time dependent baryonic parameters.}  
\resizebox{\columnwidth}{!}{%
\begin{tabular}{ | l | l| l | l | l | l | l | l | }
\hline  
$\nu_i$ : & $\log_{10} M_{\rm c}$ & $\mu$ & $\theta_{\rm ej}$  & $\gamma$  & $\delta$ & $\eta$ & $\eta_{\delta}$  \\ \hline 
{\bf Lower} & -0.02 &  -0.05 & -0.04   & -0.04  &  -0.04  & -0.05  & -0.05  \\
{\bf Upper} & 0.015  &  1.   &  0.146  & 0.146  &  0.131   & 0.296  & 0.296 \\ 
{\bf Fiducial} & 0.01  & 0  & 0  &0  &  0  & 0  &  0.06 \\ 
\hline
\end{tabular}
}\label{tab:all_params_nu}
\end{table}

\subsection{Baryons: \bcemu{}} 
\label{sec:baryons}
% Importance of baryons
Beyond massive neutrinos and modified gravity, we must also consider the effects of baryons as we move into the nonlinear regime. These have been shown to have a significant impact on the matter power spectrum and the cosmic shear spectrum \citep{Semboloni:2011fe,Mead:2020vgs,Euclid:2020tff,Schneider:2019snl,Schneider:2019xpf,Arico:2020lhq}. 

% BCemu parameterisations
\bcemu{} \citep[\href{https://github.com/sambit-giri/BCemu}{\faicon{github}}]{Giri:2021qin} is a state-of-the-art baryonic boost, $B^{\rm baryons}$, emulator giving the nonlinear effects of baryons on the matter power spectrum. It requires a set of cosmology-independent parameters as input, augmented with the baryon fraction $f_{\mathrm{b}} = \Omega_{\rm{b}} / \Omega_{\rm{m}}$. Some \bcemu{} parameters have also been shown to have a time evolution to varying degrees, depending on which hydrodynamical simulation they are fit to. In \cite{Giri:2021qin} the authors propose a power law dependence 
\begin{align}
\theta_i(z) = \theta_{i,0} \, a^{\nu_i} \, ,
    \label{eq:baryonevol}
\end{align}
where $\nu_i$ is a free parameter giving the time dependence of the \bcemu{} baryon input parameter $\theta_i$, with the value today, $\theta_{i,0} = \theta_i(a=1)$, also being a free parameter. \cite{Giri:2021qin} investigate two parameterisations:
\begin{itemize}
    \item \mbox{a 7-parameter model, $\theta_i \in \{\log_{10}M_{\rm c},  \eta_\delta, \theta_{\rm ej}, \mu, \gamma, \delta, \eta \}$};
    \item a 3-parameter model,  $\theta_i \in \{\log_{10}M_{\rm c},  \eta_\delta, \theta_{\rm ej}\}$. 
\end{itemize}
We refer to \citet{Giri:2021qin} for an in-depth discussion of the meaning of each of these parameters. We note that the 3-parameter model is motivated by its ability to fit within $1\%$ a large suite of hydrodynamical simulation measurements at a fixed redshift. One of our goals (see \autoref{sec:forecasts}) is to investigate to what degree this reduced parameterisation improves or biases cosmological inference. Note that an $N$-parameter model actually has $N\times 2$ degrees of freedom when including the time dependence. 

% BCemu ranges and fiducial
In \autoref{tab:all_params} we list the chosen \bcemu{} parameters along with their fiducial values today ($a=1$) and prior ranges that we select for our analysis. The ranges are $20\%$ tighter than the defaults implemented in the \bcemu{} software. The reason for this is to prevent any deviations outside the \bcemu{} allowed ranges when we consider higher redshifts, given the time dependence we impose. These ranges are also the edges of the uniform prior distributions used for these baryon parameters in our Markov Chain Monte Carlo (MCMC) analyses of \autoref{sec:forecasts}. Priors on the power law parameters $\nu_i$ are given in \autoref{tab:all_params_nu}, which are also chosen so as not to exceed the allowed \bcemu{} range at any redshift considered in this work ($0\leq z \leq 5$). As fiducial values for these parameters, we take $\nu_{\log_{10} M_c} =0.01 $ and $\nu_{\eta_\delta}=0.06$,  with all other $\nu_i=0$. 

\subsection{Cosmic Shear}
% nonlinear matter power spectrum
We model the nonlinear matter power spectrum $P_{\rm NL}$ as
\begin{equation}
    P_{\rm NL} = B^{f(R) + \nu}_{\rm REACTEMU-FR} \times B^{\rm baryons}_{\rm BCEMU} \times P_{\rm HMCODE}^{\rm \Lambda CDM} \, ,
    \label{eq:datavector}
\end{equation}
where nonlinear LCDM effects are modelled using a \cosmopower{} emulator of \HMcode{} \citep{Mead:2016zqy}, the baryonic effects are modelled using \bcemu{} and the massive neutrino and $f(R)$-gravity effects by \reactemufr{}. Note that the \HMcode{} emulator as well as the \bcemu{} $f_{\rm b} = \Omega_{\rm b}/\Omega_{\rm m}$ parameter completely overlap the ranges taken for \reactemufr{} (see \autoref{tab:all_params}).

% pure shear power
The cosmic shear power spectrum $C_{ij}^{\gamma \gamma}(\ell)$ can be obtained from the nonlinear matter power spectrum $P_{\mathrm{NL}}(k, a)$ as an integral along the line of sight \citep[\textit{e.g.}][]{Bartelmann_2001, Kilbinger_2015, Kilbinger_2017, Kitching_2017, Asgari_2021}:
\begin{align}
\label{eq:cosmic_shear}
    C_{ij}^{\gamma \gamma}(\ell) = \int_0^{\chi_{\mathrm{H}}} d\chi \frac{W^{\gamma}_{i}(\chi)W^{\gamma}_{j}(\chi)}{\chi^2} P_{\rm NL}\left(\frac{\ell+1/2}{\chi}, a\right) \, ,  
\end{align}
where $\chi$ is the comoving distance as a function of the scale factor, $\chi_{\mathrm{H}}$ is the Hubble radius $\chi_{\mathrm{H}}=c/H_0$, and $W^{\gamma}_{i}$ are weighting functions for each redshift bin $i$,
\begin{align}
W^{\gamma}_{i} (\chi) &= \frac{3 \, H_0^2 \, \Omega_{\mathrm{m}}}{2\, c^2}
\frac{\chi}{a} \int_{\chi}^{\chi_{\mathrm{H}}} \mathrm{d} \chi'\; n_{i}(\chi')\; \frac{\chi' - \chi}{\chi'}\; \ ,
\end{align}
where $n_i$ is the redshift distribution for bin $i$, $H_0$ is the Hubble constant and $\Omega_{\mathrm{m}}$ is the total matter density parameter. Note that throughout the paper we assume the extended Limber approximation \citep{Kaiser_1998,LoVerde_2008} for the projected power spectra.

% IA power
The lensing signal is contaminated by the intrinsic alignment (IA) of galaxies. Similarly to \citet{SpurioMancini:2021ppk} and \citet{ManciniSpurio:2021jvx}, we model the IA contamination adding a sum of IA contributions to \autoref{eq:cosmic_shear}, namely
\begin{align}
    C^{\gamma \mathrm{I}}_{ij}(\ell) + C^{\mathrm{I}\gamma}_{ij}(\ell) + C^{\mathrm{II}}_{ij}(\ell) \ ,
\end{align}
where the superscript $\mathrm{I}$ refers to the IA weighting function
\begin{align}
\label{eq:w_ia}
W^{\rm I}_{i} (\chi) &= - A_{\rm IA}\; \left( \frac{1+z}{1+z_{\rm pivot}} \right)^{\eta_{\rm IA}}  \frac{C_1 \, \rho_{\rm cr} \, \Omega_{\rm m}}{D(\chi)}\; n_{i}(\chi)\;.
\end{align}
$D(\chi)$ in \autoref{eq:w_ia} represents the linear growth factor, $\rho_{\mathrm{cr}}$ denotes the critical density, $C_1$ is a constant, and $z_{\mathrm{pivot}}$ is an arbitrary redshift that has been set to 0.3. The IA contamination is parameterised by an amplitude $A_{\rm IA}$ and a power law redshift dependence parameter $\eta_{\rm IA}$ \citep[see \textit{e.g.}][]{Bridle:2007ft}. We remark that intrinsic alignment modelling and priors are also an open question \citep[see][for a recent study]{Samuroff:2020gpm}. We defer this issue to future work and focus on the impact of baryonic, massive neutrino and beyond-LCDM modelling. 

\begin{figure*}
    \centering   
    \includegraphics[width=0.45\textwidth, height=0.6\textheight]{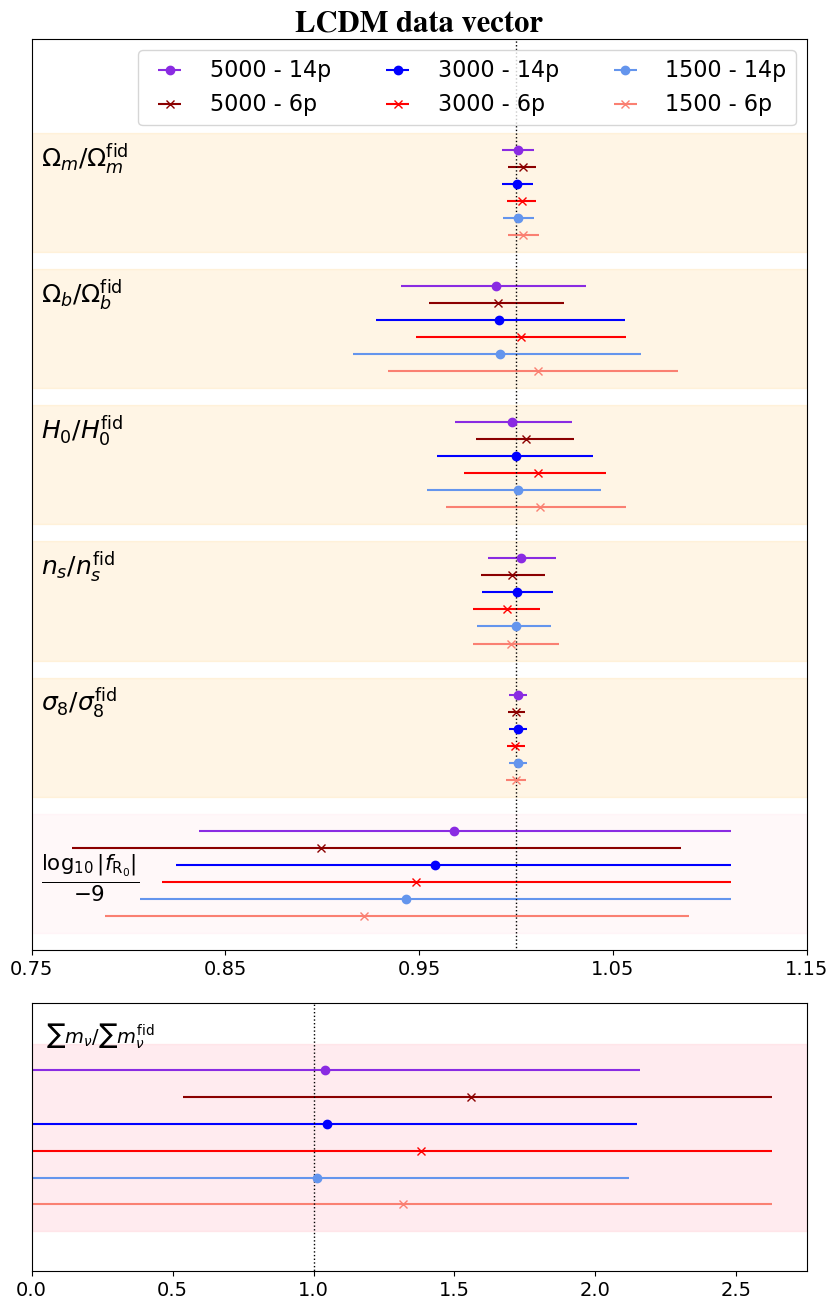}
    \includegraphics[width=0.45\textwidth, height=0.6\textheight]{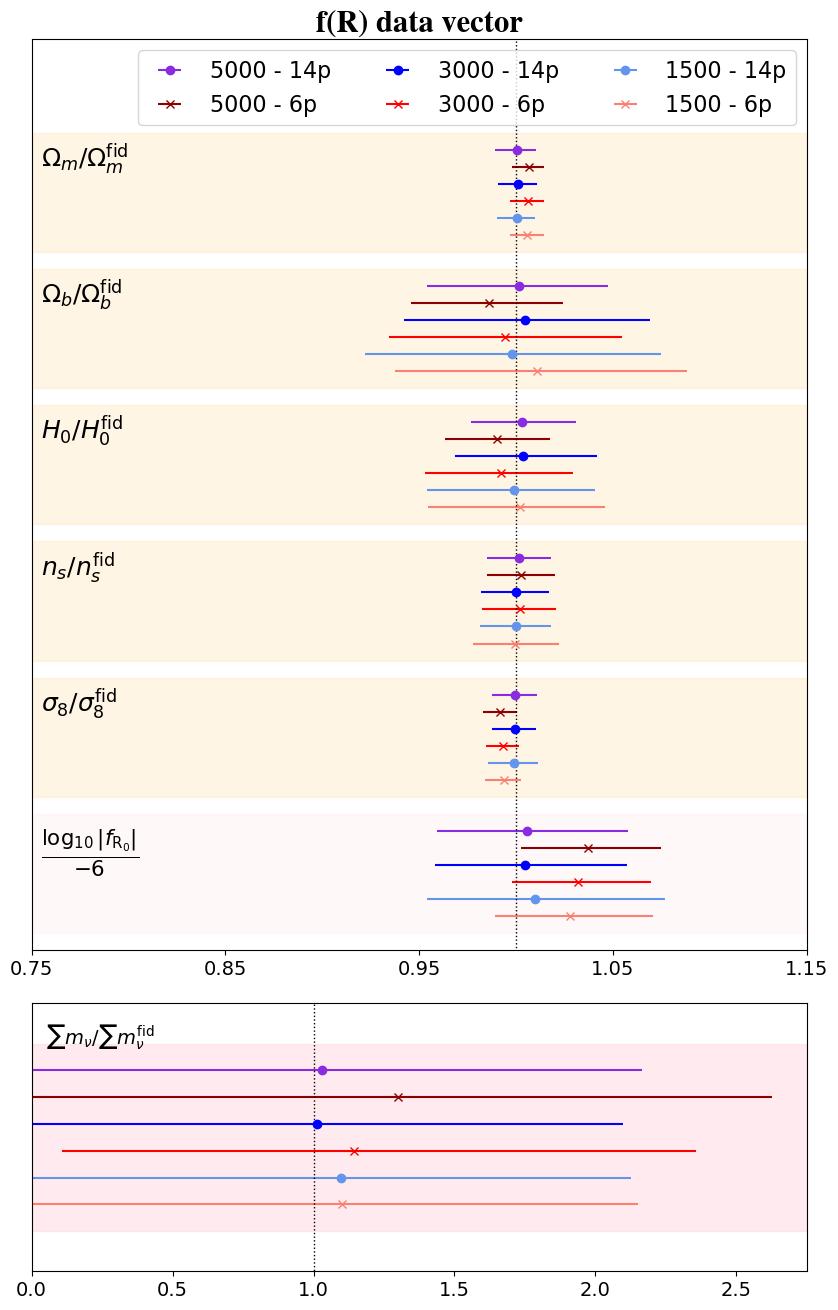}
    \caption{The marginalised means and 2$\sigma$ constraints on the cosmological and beyond-LCDM parameters, normalised to their fiducial values for Scenario A ({\bf left}) and Scenario B ({\bf right}), and for all scale cuts and baryonic parameter models.}
    \label{fig:par_shift}
\end{figure*}
\begin{table*}
  \renewcommand{\arraystretch}{2}
  \setlength{\tabcolsep}{3.5pt}
\centering
\caption{Percent 2$\sigma$ constraints on cosmological and baryonic parameters (at $z=0$) for both Scenarios, both baryonic parameter models and all scale cuts. For Scenario A, we quote the upper $2\sigma$ bound for $\log_{10}|f_{\rm R_0}|$ as the fiducial is $f_{\rm R_0}=0$. The percent constraints are calculated as $\Delta 2 \sigma / \bar{\theta}  \times  100 \%$, where $\Delta 2 \sigma$ is the $2\sigma$ confidence region (upper minus lower) and $\bar{\theta}$ is the marginalised mean parameter value.} 
\begin{tabular}{ | l | l | l| l | l | l | l | l | l | l | l | l | l | l | l | l | l | }
\hline  
 & Model &$\ell_{\rm max}$ & $\Omega_{\rm m}$  & $\Omega_{\rm b}$  & $h$ & $n_{\rm s}$ & $\sigma_8$  & $\sum m_\nu$ & $ \log_{10}|f_{\rm R_0}|$ & $\log_{10} M_{\rm c}$ & $\mu$ & $\theta_{\rm ej}$  & $\gamma$  & $\delta$ & $\eta$ & $\eta_{\delta}$  \\ \hline 
\multirow{9}{*}{{\bf A}} & \multirow{3}{*}{{\bf 14p}} 
& 1500   & 1.6 & 15.0 & 9.0 & 3.8 & 0.9 & 233.8 & -7.25 [upper] & 5.8 & 85.4 & 66.5 & 63.5 & 44.7 & 146.3 & 179.2 \\
 
  & & 3000  & 1.6 & 13.0 & 8.0 & 3.7 & 0.9 & 204.8 & -7.42 [upper] & 5.2 & 53.7 & 52.5 & 49.3 & 33.3 & 104.7 & 108.5 \\

& & 5000     & 1.7 & 9.6 & 6.0 & 3.5 & 0.9 & 222.3 & -7.53 [upper] & 3.2 & 55.8 & 33.2 & 45.6 & 21.4 & 79.0 & 99.6 \\

& \multirow{3}{*}{{\bf 6p}}  
& 1500    & 1.6 & 14.8 & 9.2 & 4.5 & 1.0 & 199.1 & -7.09 [upper] & 3.0 & - & 21.8 & - & - & - & 88.6 \\
 
& & 3000    & 1.5 & 10.8 & 7.3 & 3.5 & 0.9 & 190.1 & -7.35 [upper] & 2.3 & - & 14.4 & - & - & - & 85.0 \\

& & 5000     & 1.4 & 7.0 & 5.0 & 3.3 & 0.9 & 134.1 & -6.94 [upper] & 1.7 & - & 9.8 & - & - & - & 82.4 \\

& \multirow{3}{*}{{\bf fixed}}  
& 1500      & 1.1 & 9.6 & 7.1 & 3.1 & 0.8 & 225.0 & -7.98 [upper]& - & - & - & - & - & - & -\\

& & 3000    & 0.9 & 5.7 & 5.5 & 3.2 & 0.6 & 166.1 & -8.28 [upper]& - & - & - & - & - & - & - \\ 
 
& & 5000    & 0.7 & 4.8 & 4.3 & 3.1 & 0.5 & 104.6 & -8.37 [upper]& - & - & -& - & - & - & -\\ \hline

  \multirow{6}{*}{{\bf B}} & \multirow{3}{*}{{\bf 14p}}  
& 1500       & 2.0 & 15.3 & 8.7 & 3.7 & 2.6 & 193.6 & 12.2 & 6.0 & 89.8 & 63.4 & 73.4 & 55.0 & 119.2 & 147.0 \\

& & 3000     & 2.0 & 12.6 & 7.3 & 3.5 & 2.3 & 207.0 & 9.9 & 5.4 & 58.4 & 45.3 & 53.0 & 37.3 & 98.0 & 104.6  \\

& & 5000     & 2.1 & 9.3 & 5.4 & 3.3 & 2.3 & 221.7 & 9.8 & 3.2 & 49.4 & 35.0 & 43.7 & 32.0 & 79.8 & 102.9 \\ 

& \multirow{3}{*}{{\bf 6p}}  
& 1500       & 1.8 & 14.9 & 9.1 & 4.5 & 1.9 & 195.8 & 8.0 & 2.8 & - & 24.5 & - & - & - & 86.8 \\

& & 3000     & 1.7 & 12.1 & 7.7 & 3.8 & 1.8 & 196.5 & 6.9 & 2.0 & - & 17.3 & - & - & - & 84.1 \\

& & 5000     & 1.6 & 7.9 & 5.5 & 3.5 & 1.8 & 229.5 & 7.0 & 1.2 & - & 12.1 & - & - & - & 83.4 \\ 

& \multirow{3}{*}{{\bf fixed}}  
& 1500       & 1.5 & 8.3 & 5.4 & 3.5 & 1.6 & 221.2 & 4.6 & - & - & - & - & - & - & -\\

& & 3000     & 1.4 & 6.2 & 4.6 & 3.3 & 1.4 & 179.6 & 4.1 & - & - & - & - & - & - & - \\

& & 5000     & 1.0 & 5.2 & 4.1 & 3.2 & 1.1 & 126.8 & 3.6 & - & - & -& - & - & - & -\\ \hline 

\end{tabular}
\label{tab:constraints}
\end{table*}

% ================================================
%                   RESULTS
% ================================================
\section{Forecasts for Stage IV surveys}
\label{sec:forecasts}
\subsection{Setup}
% specs
Our mock survey configuration has the following Stage IV specifications: sky fraction $f_{\rm sky} = 0.4$, number density of galaxies per arcminute squared $n = 30\, \mathrm{arcmin}^{-2}$ and shape noise parameter $\sigma_e = 0.3$. We take 10 equipopulated tomographic bins with galaxies following $n_{\rm i}(z) \sim z^2 \exp(-(z/z_0)^{3/2})$ \citep{Smail_1995,Joachimi_2010}, with $z_0=0.9$. For each redshift bin $i = 1, \dots 10$ we include a parameter $D_{z_{\rm i}}$ that shifts the mean of the redshift distribution in the bin \citep{Asgari_2021,SpurioMancini:2021ppk}. We vary these parameters assuming a Gaussian prior $\mathcal{N}(0, 10^{-4})$ for each of them, following \citet{SpurioMancini:2021ppk}.

% 2 scenarios
We perform MCMC analyses on a set of mock cosmic shear data vectors, adopting the fiducial cosmology and baryonic boost parameters given in \autoref{tab:all_params}. We then consider two scenarios:
\begin{enumerate}
    \item[\bf (A)] 
    \emph{The Universe is LCDM} ($|f_{\rm R_0}|=0$). We aim at checking to what extent Stage IV experiments can constrain $f_{\rm R_0}$ and $\sum m_\nu$. 
    \item[\bf (B)] 
    \emph{The Universe is $f(R)$} ($|f_{\rm R_0}|=10^{-6}$). We aim at checking to what extent Stage IV experiments can detect $|f_{\rm R_0}|>0$.
\end{enumerate} 
In all analyses we vary all 5 cosmological parameters as well as $|f_{\rm R_0}|$ \& $\sum m_{\nu}$. To investigate the importance of the baryon parameters when looking to constrain cosmology and gravity, we perform analyses varying both the $7\times2$- and  $3\times2$-parameter sets (see \autoref{sec:baryons}). Note that all fiducial spectra are generated using the $7\times2$-parameter set. Lastly, to examine the impact that nonlinear scales have on the degeneracies, biases and constraints of the various parameters, we adopt three scale cuts, $\ell_{\rm max} = 1500$ (pessimistic), $\ell_{\rm max} = 3000$ (realistic) and $\ell_{\rm max} = 5000$ (optimistic). The pessimistic case is well within the current capabilities of \react{} and available fitting formulae such as \HMcode{}, as shown in \cite{Bose:2020wch}, which is what we employ here. The realistic case is also possibly within the current capabilities of available codes, for example using \react{} to construct the boost with the newest version of \HMcode{} \citep{Mead:2020vgs}, and combining with state-of-the-art LCDM emulators such as the Euclid Emulator \citep{Euclid:2020rfv}. Finally, the optimistic case represents a possible modelling potential given, for example, high quality emulators for the pseudo power spectrum \citep{Giblin:2019iit}. To implement these choices we use 50 $\ell$ bins in the range $[30,5000]$, which for our $\ell_{\rm max}=1500, 3000$ runs we cut to multipoles up to the corresponding values. For each setup we assume a simple Gaussian covariance matrix \citep{Joachimi_2007}. Note that we leave for future work the inclusion of non Gaussian \citep{Sato_2013,Krause_2017} and super sample \citep[][]{Takada_2013,Li_2014,Lacasa_2018} contributions (see also a discussion on their possible impact in \autoref{sec:conclusions}).

In summary, we study 3 scale cuts and 2 baryonic models for each of the 2 scenarios, giving a total 12 analyses. We vary 33 (25) parameters in total: 5 cosmological $+ \sum m_\nu$ \& $\log_{\rm 10}|f_{\rm R_0}|$, 14 (6) baryonic, 2 intrinsic alignment and 10 redshift distribution shift parameters. We sample our posterior distribution using \textsc{PolyChord} \citep{Handley_2015a,Handley_2015b}, using 1000 live points to ensure accurate contours despite the large parameter space. A single likelihood evaluation takes approximately 0.7 seconds on a single Intel Xeon E5640 CPU core. \textsc{PolyChord} does not allow one to specify the number of samples as the sampling terminates once a given accuracy threshold is reached. This takes longer for more complex posteriors, i.e. for a larger number of parameters and a larger scale cut. Our longest run, the 14 parameter, $\ell_{\rm max}=5000$, run took approximately one day on 512 cores.

%===================================================
\subsection{Results}
The $2\sigma$ constraints on all cosmological and baryonic parameters are summarised in \autoref{tab:constraints} for all cases considered in this work. We also show the constraints and shifts of the recovered cosmological parameters in \autoref{fig:par_shift}. Note that $\log_{10}|f_{\rm R_0}|$ takes values below the effective LCDM-limit of $-9$ as we emulate down to $-10$. In \autoref{app:fullpost} we give the full set of posterior distributions for all parameters except the redshift-dependency parameters of the baryonic effects, which we find to be largely unconstrained. This is provided for both scenarios, all scale cuts and both choices of baryonic parameter sets. 

% scenario A results
\subsubsection{Scenario A}

\begin{figure*}
    \centering
    \includegraphics[width=\textwidth, height=0.95\textheight]{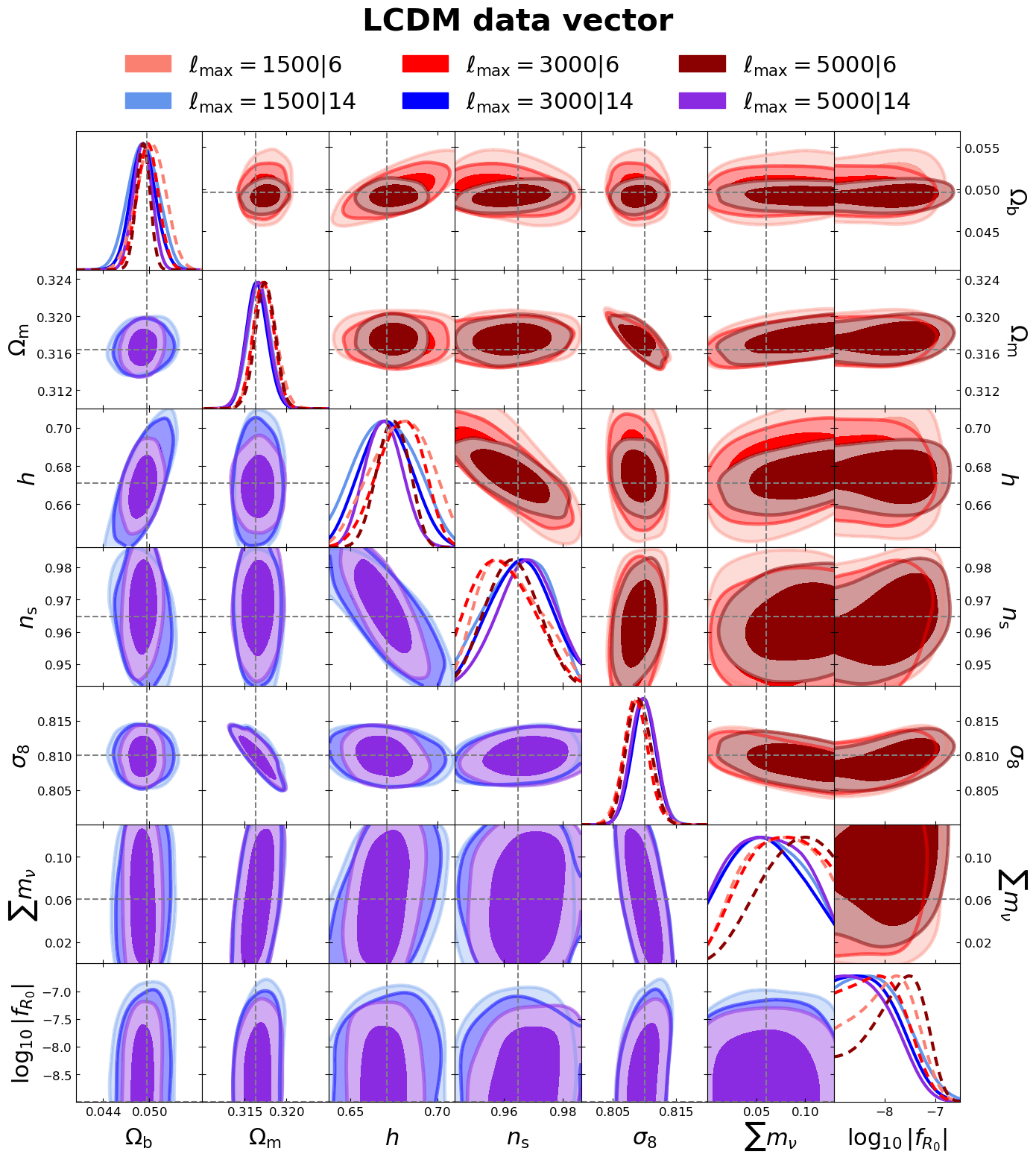}
    \caption{2-dimensional cosmological parameter posteriors for the $7\times2$ ({\bf bottom-left}) and $3\times2$ ({\bf top-right}) baryonic parameter models in {\bf Scenario A} (LCDM data vector). We show the $\ell_{\rm max}=1500,3000,5000$ cases for both baryonic models.}
    \label{fig:cosmo_params_lcdm}
\end{figure*}

% cosmological parameters
In \autoref{fig:cosmo_params_lcdm} we show the marginalised 2-dimensional 95\% ($2\sigma$) and 68\% ($1\sigma$) distributions for the cosmological parameters as well as the beyond-LCDM parameters $\sum m_\nu$ and $|f_{\rm R_0}|$. We summarise the main findings below and forward the reader to \autoref{tab:constraints} for the exact constraints.
\begin{itemize}
    \item 
    In the conservative case ($\ell_{\rm max}=1500$) we find that cosmic shear is able to constrain all LCDM cosmological parameters at least at the $\sim 10\%$ level (95 \% C.L.), with almost an order of magnitude better constraints on $\Omega_{\mathrm{m}}$ and $\sigma_8$.  $n_{\mathrm{s}}$ is also better constrained, but to a lesser extent ($3.8\%$).
    \item 
    The inclusion of scales $1500 \leq \ell \leq 5000$ does not significantly improve the base LCDM cosmological parameters, with the most significant improvement being in constraints on $h$ (up to a 6\% measurement) and $\Omega_{\mathrm{b}}$ (roughly a factor of 2 with respect to the conservative case).
    \item 
    Even in the conservative case ($\ell_{\rm max}=1500$) cosmic shear is able to constrain $|f_{\rm R_0}| \leq 10^{-7}$ (95 \% C.L.), while $\sum m_\nu$ remains largely unconstrained. 
    \item 
    The inclusion of scales $1500 \leq \ell \leq 5000$ does not significantly improve the constraints on $\sum m_\nu$ and $|f_{\rm R_0}|$.
    \item 
    Reducing the number of baryonic degrees of freedom from $7\times2$ to $3\times2$ does not notably improve constraints on any parameter nor does it lead to any significant (up to $1\sigma$) biases in the recovered cosmological parameters. We find at most $\sim 1 \sigma$ shifts on the $\sum m_{\nu}$ and $|f_{\rm R_0}|$ marginalised posteriors in the $\ell_{\rm max}=5000$ case. 
\end{itemize}

% Baryons dampen cosmological information
These results show that the highly nonlinear scales have only a marginal amount of information on massive neutrinos and modified gravity which is not degenerate with baryonic degrees of freedom. Similarly, the baryonic degrees of freedom dampen the cosmological information one is able to extract from small, nonlinear scales. This was already shown in \cite{Copeland:2017hzu}. There the authors illustrate how cosmological and dark energy constraints degrade with marginalisation of baryonic effects and that there is little improvement by going to smaller scales. Further, highly consistent results were also found in \cite{Bose:2020wch}, where similar poor gains in constraining power on $\log_{10}|f_{\rm R_0}|$ was found when including scales $1500 \leq \ell \leq 3000$. They do however obtain worse overall constraints compared to ours, by almost half an order of magnitude on $f_{\rm R_0}$, despite their analysis not including baryonic degrees of freedom nor massive neutrinos. This is likely due to the difference in setup, as for example we use 10 tomographic bins while they only considered 3.

\begin{turnpage}
\begin{figure*}
    \centering
    \includegraphics[height=0.9\textwidth, width=0.59\textheight]{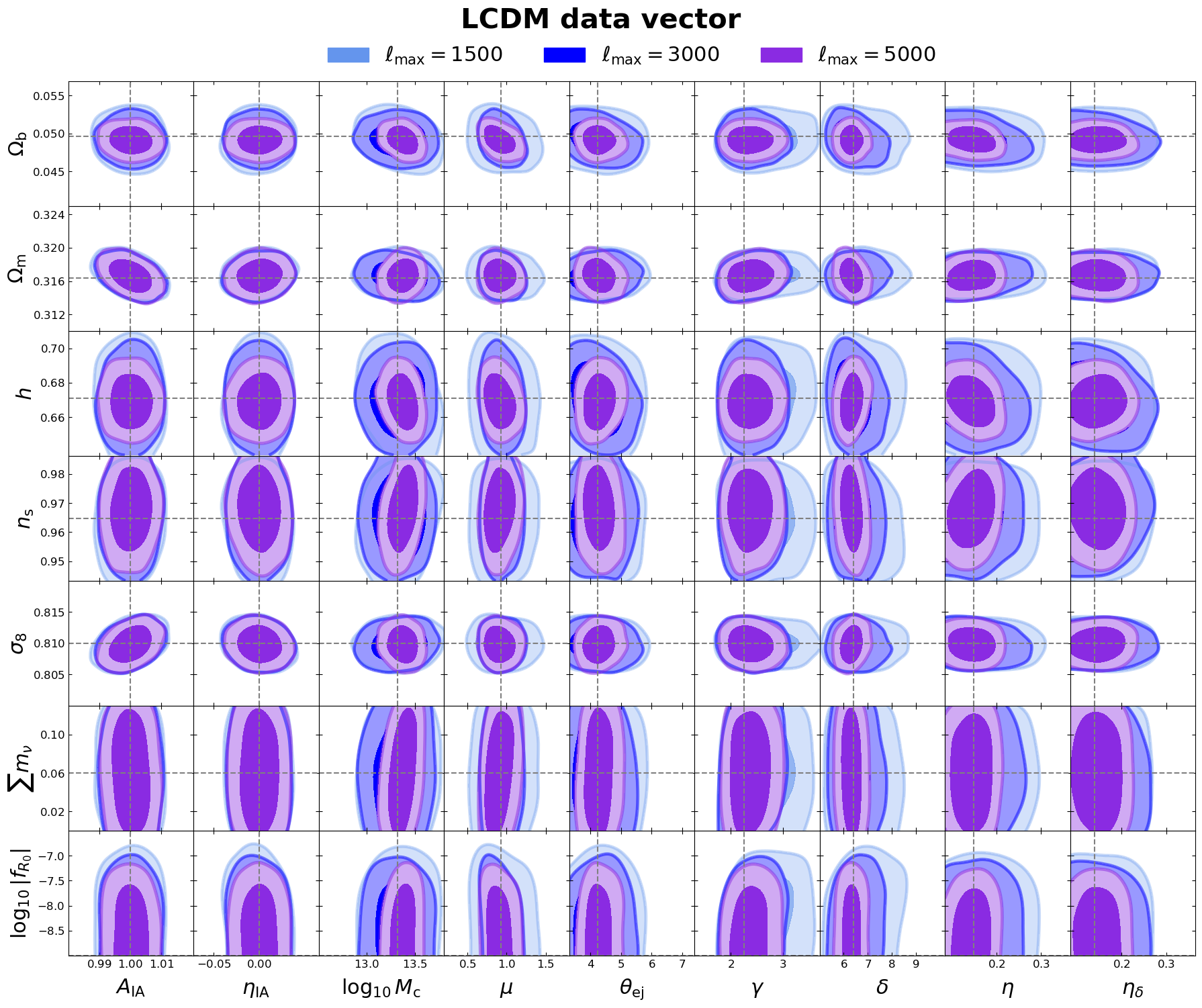}
    \includegraphics[height=0.87\textwidth, width=0.4\textheight]{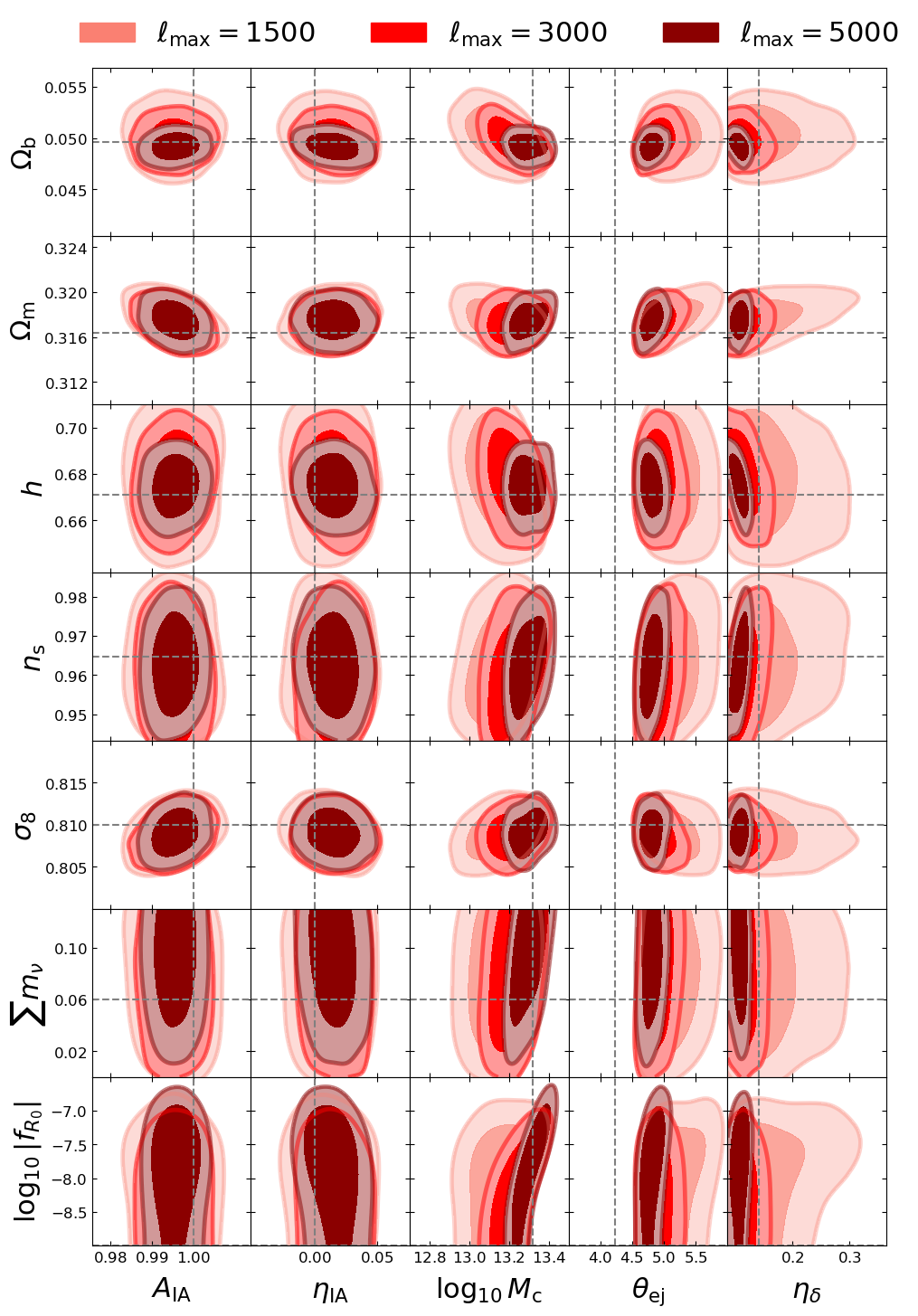}
    \caption{2-dimensional posteriors between cosmological parameters and baryonic parameters at $z=0$ for the $7\times2$ ({\bf left}) and $3\times2$ ({\bf right}) parameter models in {\bf Scenario A} (LCDM data vector). We show the $\ell_{\rm max}=1500,3000,5000$ cases for both baryonic models.}
    \label{fig:nuisance_params_lcdm}
\end{figure*}
\end{turnpage}

% Nuisance parameters / cosmological parameters correlation
To investigate further the origin of this lack of constraining power on cosmological parameters at highly nonlinear scales, we plot the posteriors for the cosmological, intrinsic alignment and baryonic degrees of freedom. This is shown in \autoref{fig:nuisance_params_lcdm}. It is clear from these plots that much of the information at nonlinear scales is associated with the baryonic degrees of freedom, with all of these parameters seeing significant gains in constraints when moving from $\ell_{\rm max}=1500$ to $\ell_{\rm max}=5000$. This is also clearly seen in \autoref{tab:constraints}. Notably, in the $7\times 2$ parameter case, there are no large degeneracies between baryonic degrees of freedom and cosmological ones, which supports the lack of improvement in constraining power on cosmological parameters when moving to nonlinear scales. Baryonic parameters are in general less constrained than cosmological parameters, with the exception of $\log_{10} M_{\mathrm{c}}$ which can be constrained down to the $\sim 5$\% level even in conservative configurations. However, baryonic parameters are in general more heavily biased, up to $2 \sigma$ for $\theta_{\mathrm{ej}}$, when using the reduced model with $3 \times 2$ parameters over the one with $7 \times 2$ parameters.

\begin{figure*}
    \centering
    \includegraphics[width=\textwidth, height=0.8\textheight]{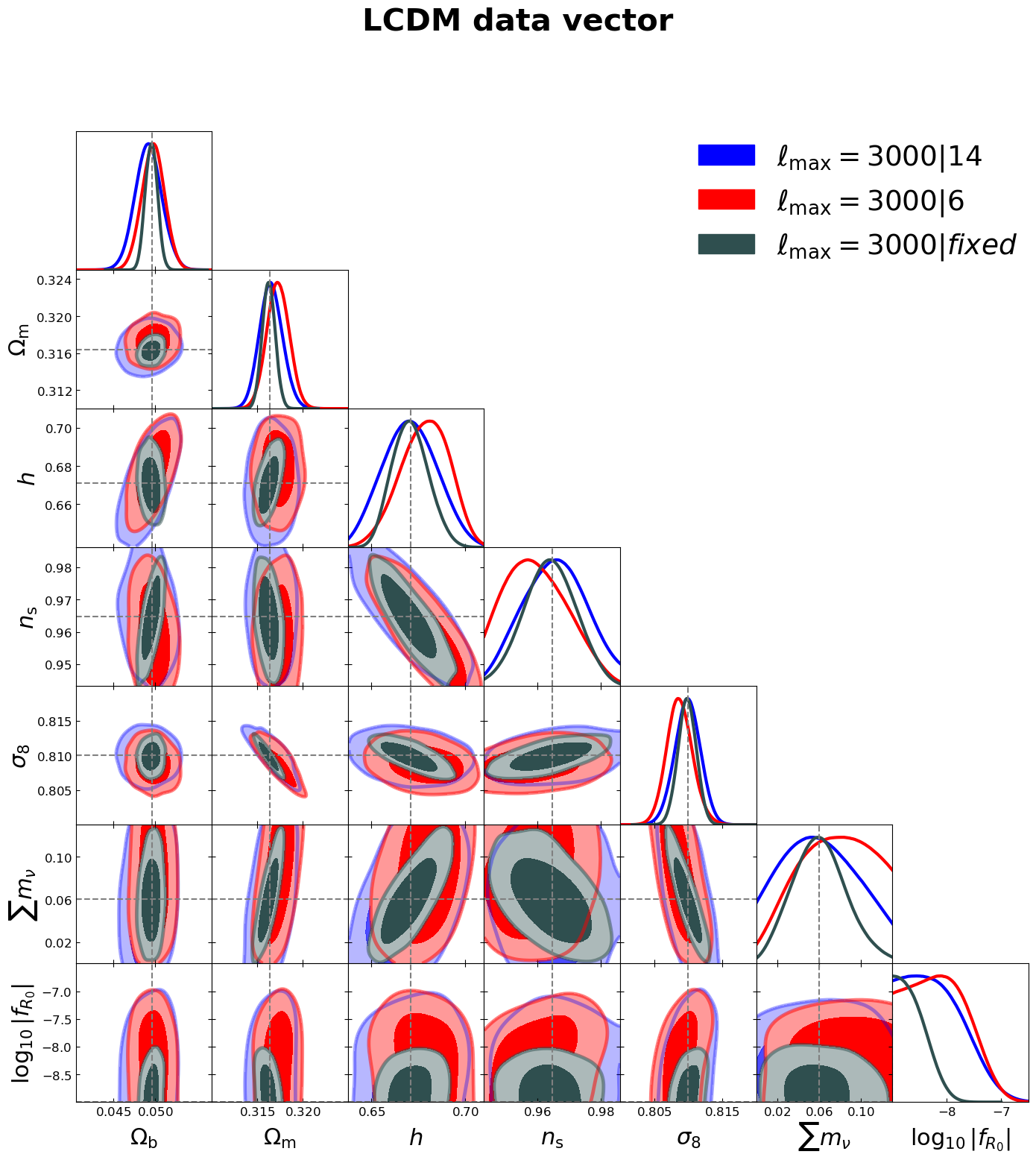}
    \caption{Selected 2-dimensional cosmological parameter posteriors for analyses varying $7\times2$ (blue), $3\times2$ (red) and no (green) baryonic parameters in {\bf Scenario A} (LCDM data vector) using $\ell_{\rm max}=3000$.}
    \label{fig:fixed_baryon_lcdm}
\end{figure*}

% Fixing baryons
In \autoref{tab:constraints} we also show the marginalised constraints for {\bf Scenario A} analyses but with all baryonic degrees of freedom fixed to their fiducial values. As in \cite{Schneider:2019snl}, this represents an ideal situation where baryonic effects are completely known \footnote{One can constrain these degrees of freedom from other observations such as gas X-ray emissions \citep{eROSITA:2012lfj, Schneider:2019xpf}.}. In \autoref{fig:fixed_baryon_lcdm} we plot the posterior contours for the cosmological parameters in this setup where we assume perfect knowledge of the baryonic parameters, for the realistic configuration $\ell_{\mathrm{max}} = 3000$, and we compare the results against those obtained with the same $\ell$-cut, but varying either all of the $7 \times 2$ or all of the $3 \times 2$ baryonic parameters. We see that knowing the baryonic effects improves constraining power, relative to the $7\times 2$ parameter model, on $\Omega_m$, $h$ and $\sum m_\nu$ by 50\%, and by almost an order of magnitude on $f_{\rm R_0}$. Unlike \cite{Schneider:2019snl} we see little improvement on $n_s$, which is likely due to the mild degeneracy of $n_s$ with $f_{\rm R_0}$ and $\sum m_\nu$. As expected, the constraint on $\Omega_{\mathrm{b}}$  is significantly improved, by roughly factor of 2, which just indicates that the baryon fraction is degenerate with baryonic effects.

% scenario B results
\subsubsection{Scenario B}

\begin{figure*}
    \centering
    \includegraphics[width=\textwidth, height=0.95\textheight]{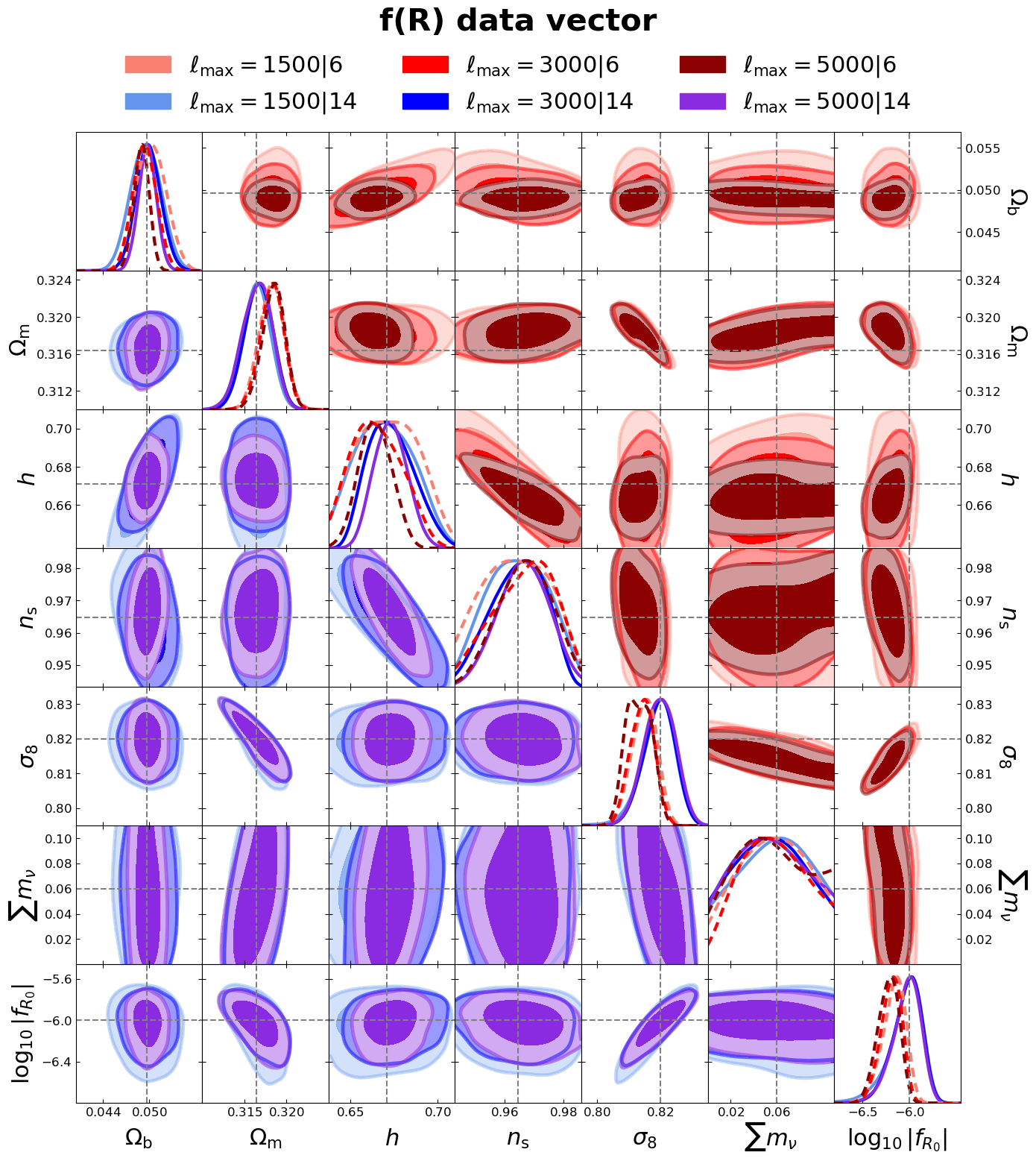}
    \caption{Same as \autoref{fig:cosmo_params_lcdm} but for {\bf Scenario B}.}
    \label{fig:cosmo_params_fr}
\end{figure*}

% degeneracies between cosmological parameters
In \autoref{fig:cosmo_params_fr} we show the marginalised 2-dimensional distributions for the cosmological parameters as well as the beyond-LCDM parameters. The main findings are consistent with {\bf Scenario A}, with marginally worse constraints on $\sigma_8$ and $\sum m_\nu$. This is driven by the strong degeneracies between $\sigma_8$, $|f_{\rm R_0}|$ and $\sum m_\nu$. At all scales, increasing $\log_{10}|f_{\rm R_0}|$ will naturally increase $\sigma_8$ as the modification works to enhance power, giving the correlation we see between these parameters. Conversely, massive neutrinos are known to suppress power and so we expect an anti-correlation between $\sum m_\nu$ and $\sigma_8$.

% No stronger constraints on f(R) from small scales
We note that weak lensing alone is able to provide a clear detection of $f(R)$ gravity, given a modification of $\log_{10}|f_{\rm R_0}|=10^{-6}$, even in the presence of massive neutrinos. This detection is not changed notably by including scales $1500 \leq \ell \leq 5000$, and is moderately improved by reducing the baryonic parameters. These results are highly consistent with the results of \cite{Bose:2020wch}, which performed a similar analysis without massive neutrinos and no baryonic effects. It is clear that even without baryons and massive neutrinos, most constraining power comes from scales $\ell \leq 1500$.

\begin{turnpage}
\begin{figure*}
    \centering
    \includegraphics[height=0.9\textwidth, width=0.59\textheight]{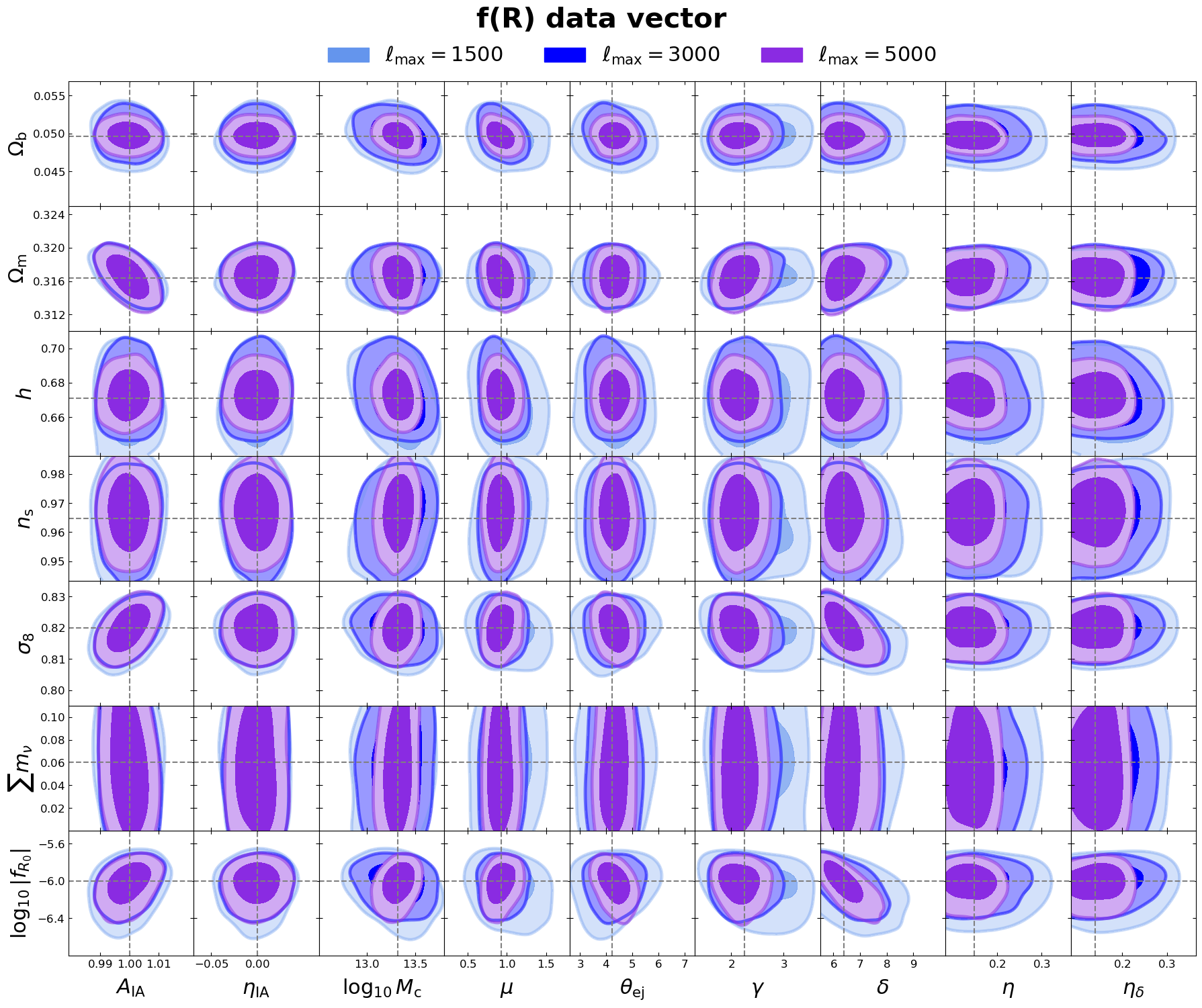} 
    \includegraphics[height=0.87\textwidth, width=0.4\textheight]{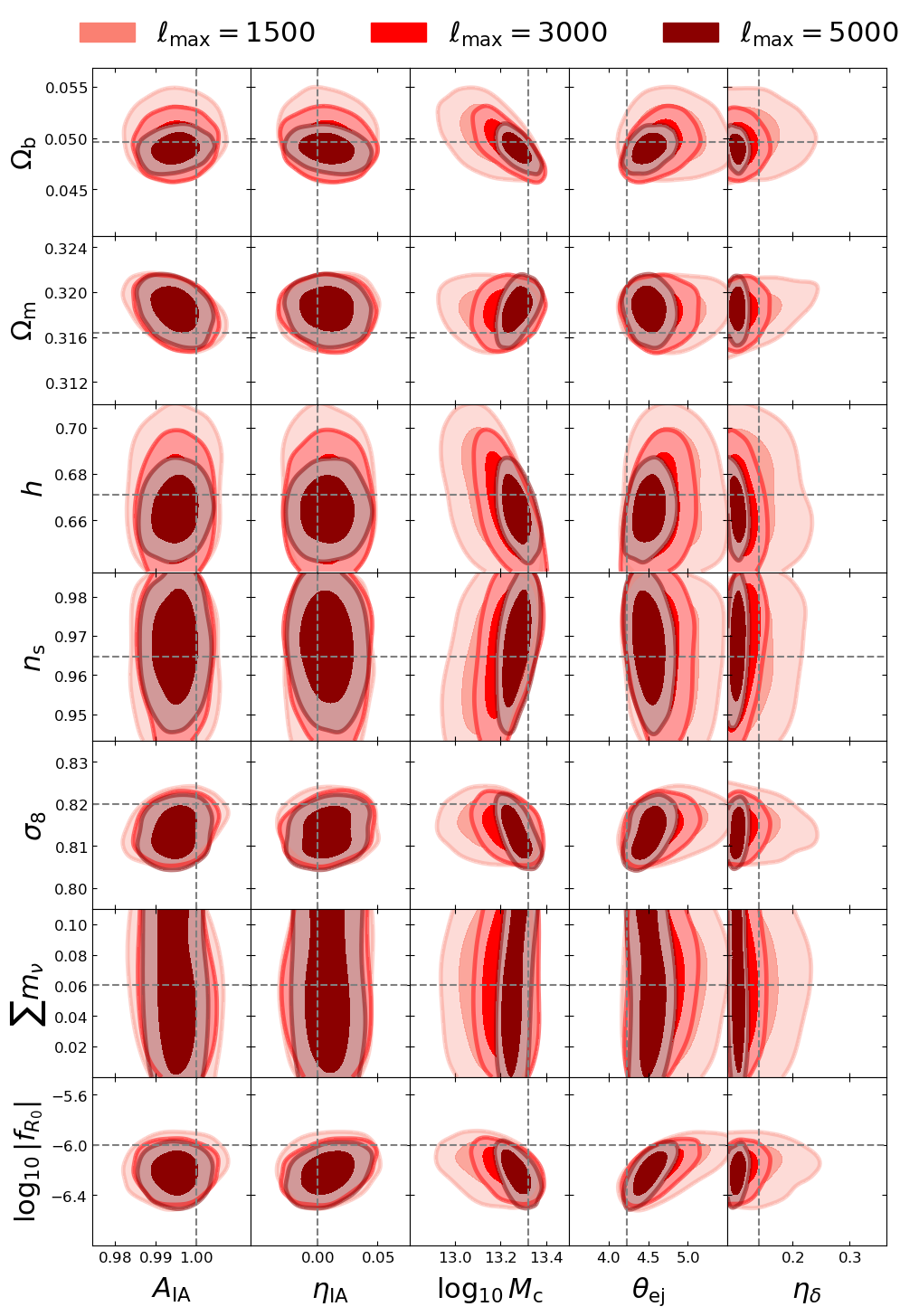}
    \caption{Same as \autoref{fig:nuisance_params_lcdm} but for {\bf Scenario B}.}
    \label{fig:nuisance_params_fr}
\end{figure*}
\end{turnpage}

% Nuisance parameters / cosmological parameters correlation
We also show the marginalised posterior distributions between the cosmological and baryonic parameters in \autoref{fig:nuisance_params_fr}. This uncovers clear degeneracies between $f(R)$ effects and baryonic physics in the reduced baryonic parameter model, which accounts for the gain in constraining power when moving to very nonlinear scales, which breaks these degeneracies. On the other hand, the full baryonic model does not exhibit such strong degeneracies and we gain less in this case when moving to highly nonlinear scales.  

\begin{figure*}
    \centering
    \includegraphics[width=\textwidth, height=0.8\textheight]{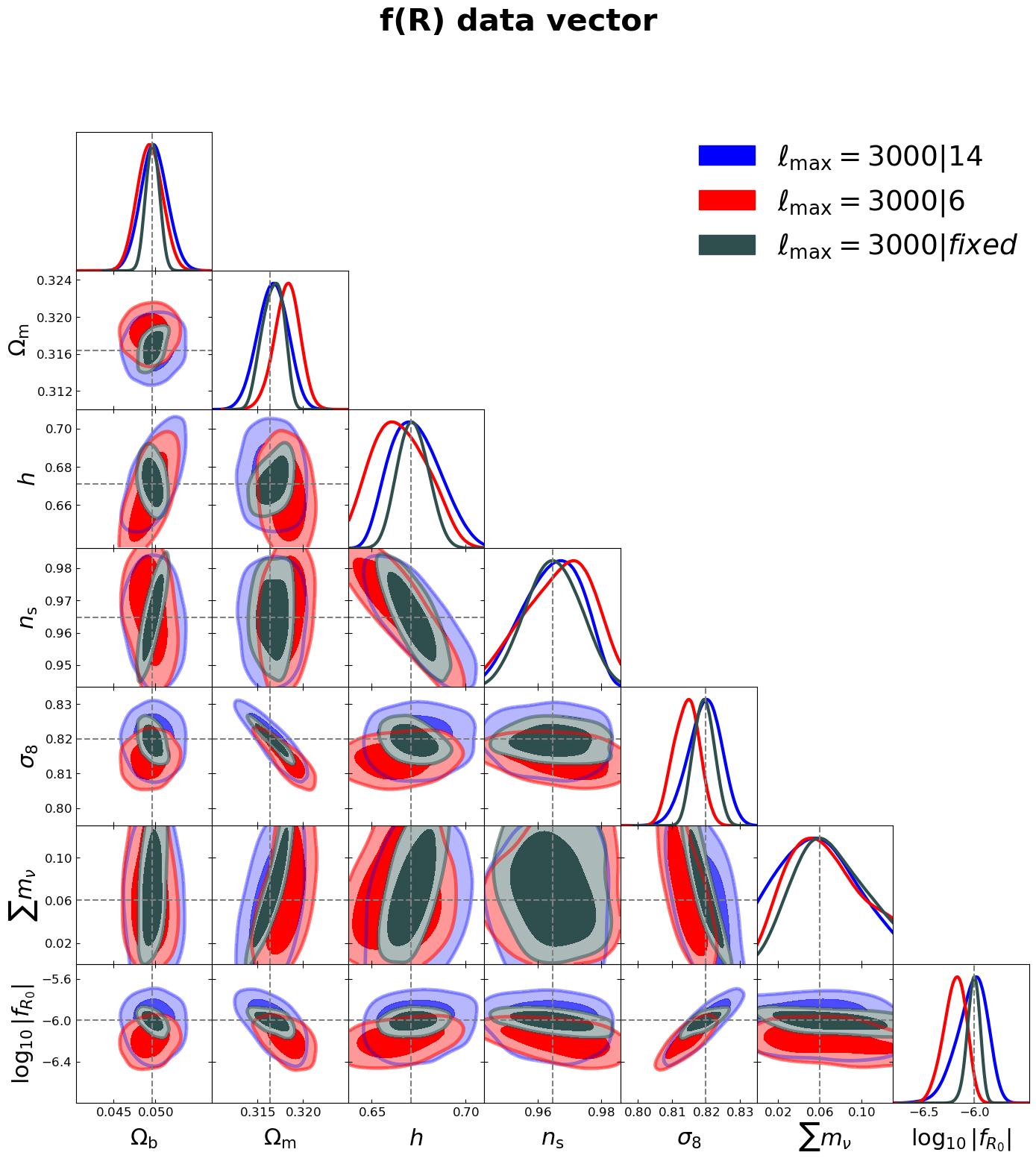}
    \caption{Selected 2-dimensional cosmological parameter posteriors for analyses varying $7\times2$ (blue), $3\times2$ (red) and no (green) baryonic parameters in {\bf Scenario B} (f(R) data vector) using $\ell_{\rm max}=3000$.}
    \label{fig:fixed_baryon_fr}
\end{figure*}

% Fixing baryons
In \autoref{tab:constraints} we report the expected constraints obtained in the realistic case $\ell_{\mathrm{max}} = 3000$ assuming perfect knowledge of the baryonic parameters. In \autoref{fig:fixed_baryon_fr} we compare these results against the contours for the $7 \times 2$ and $3 \times 2$ parameter models. From this comparison we observe that opening up the parameter space for baryonic parameters leads to significantly worse constraints on all cosmological parameters, particularly on $|f_{R_0}|$.

% ================================================
%                   Conclusions
% ================================================

\section{Conclusions}
\label{sec:conclusions}

In this work we have presented a lightening fast pipeline to perform cosmic shear analyses in beyond-LCDM scenarios. To accomplish this we have created an emulator, \reactemufr{} for the modification to the nonlinear LCDM power spectrum coming from Hu-Sawicki $f(R)$ gravity and massive neutrinos based on halo model reaction predictions. Together with the state-of-the-art baryonic feedback emulator, \bcemu{}, the pipeline has sufficient accuracy to be applied to Stage IV surveys' mildly nonlinear data (roughly multipoles $\ell_{\rm max}\leq 1500$). Combining with available tools such as the Euclid Emulator \citep{Euclid:2020rfv} would upgrade our pipeline to be safely applied to nonlinear scales, which we roughly define as multipoles $\ell \leq 3000$. Given the development of high quality emulators for the pseudo spectrum \citep{Giblin:2019iit}, a necessary component of the $f(R)$-massive neutrino boost, one could push to even larger multipoles, say $\ell \leq 5000$. We consider all these scale cuts in our paper and denote them as pessimistic, realistic and optimistic cases respectively.  

Using our pipeline, we have run a number of mock data analyses using cosmic shear, providing forecasts for forthcoming surveys on modified gravity and massive neutrinos. Our main findings are as follows: 
\begin{enumerate}
    \item 
    Scales between $1500\leq \ell \leq 5000$ offer only mild improvements to constraining power on cosmology, with most power going into constraining non-degenerate baryonic parameters.
    \item 
    Using a reduced baryonic parameter set ($3\times 2 = 6$ instead of $7\times2=14$) does not impact constraints on any cosmological parameters significantly, but does improve massive neutrino constraints marginally. Further, using this reduced model may lead to significant biases in the baryonic degrees of freedom. 
\end{enumerate}
A number of papers have provided similar forecasts in the literature. First, we compare with the LCDM forecasts of \cite{Euclid:2019clj}. We obtain surprisingly consistent constraints on $\Omega_m$, $n_{\rm s}$ and $\sigma_8$, but significantly better constraints on $\Omega_b$ and $h$, of factors of 4 and 2 respectively. The differences we observe may be caused by our higher redshift range (they stop at $z=2.5$) and our slightly higher sky fraction. Further, our nonlinear prescriptions are also very different. We make use of \HMcode{} while they employ \textsc{halofit}, and we include baryonic feedback via \bcemu{} while they do not consider feedback at all. These differences to the absolute power can have significant impacts on the constraining power \citep{Euclid:2020tff}.

In any case, it is expected that the inclusion of cross-correlations with galaxy clustering, and the full $3 \times 2$pt analysis, will give far better constraints on all cosmological parameters \citep{Euclid:2019clj}. Future surveys will also offer constraints on baryonic physics \citep{DES:2020rmk} which can be combined with other probes of baryonic physics. As we have shown, this will lead to improved constraints on massive neutrinos and $f(R)$ gravity, besides being interesting in its own right. However, we also note that our analysis has optimistically assumed a Gaussian covariance, ignoring super-sample contributions which in all likelihood will degrade the overall constraining power. A detailed investigation of this aspect is left for future work, as is the inclusion of optimal weighting schemes for physical scales \citep[\textit{e.g.}][]{Bernardeau_2014, Taylor_18, Deshpande_20}.

Moving on to studies of $f(R)$ gravity, in \cite{Bose:2020wch} they do not include baryonic or massive neutrino physics and have slightly different survey specifications, including far fewer tomographic bins (3 bins vs our 10 bins). Despite these differences, we find similar contributing constraining power when including multipoles between $1500 \leq \ell \leq 3000$ for both the LCDM fiducial and the $f(R)$ fiducial mock data analyses as well as similar overall constraints on $f(R)$. 

Similarly, \cite{Harnois-Deraps:2022bie} do not include massive neutrinos nor baryonic effects, and have significant differences in the survey specifications and setup. Notably they use 5 tomographic bins up to $z=3$, a quarter of our sky fraction, only consider auto-tomographic bins and have a restricted prior on $f_{\rm R_0}$. Given this, they naturally report weaker overall constraints in their full MCMC analyses. Interestingly, their Fisher analysis finds that the constraints on $f_{\rm R_0}$ will improve when including highly nonlinear scales, in contrast to what we find. We remind the reader though that they do not consider baryons, which also dampen the $f(R)$ constraints as well as their analysis having fixed the other cosmological parameters, and so representing an absolutely ideal scenario. 

More similar to our analysis is that of \cite{Schneider:2019snl,Schneider:2019xpf} which include a similar baryonic model as used in this work, but without redshift evolution, and including massive neutrinos and $f(R)$ gravity, albeit separately and using different theoretical prescriptions. They also use a different nonlinear power spectrum prescription, as well as only include 3 tomographic bins in  the range $0.1\leq z \leq 1.5$, and take $\ell_{\rm max}=4000$. The authors report significantly weaker constraints on cosmological, $f(R)$ and neutrino mass parameters. The number of tomographic bins may lead to our improved constraints, particularly on cosmological parameters. \cite{Schneider:2019xpf} do find almost 2 orders of magnitude worse constraints on $f_{\rm R_0}$ than \cite{Bose:2020wch} but include baryonic degrees of freedom which significantly worsen the constraints. On the other hand, we also include baryonic degrees of freedom and obtain $\sim$2.5 orders of magnitude better constraints. 

Another possibility could be the difference in tomographic binning and maximum redshift, which would contribute highly to these improved constraints as $f_{\rm R_0}$, $\sigma_8$ and $\sum m_\nu$ all have different redshift evolutions. The power spectrum becomes  highly insensitive to $f_{\rm R_0}$ at early times (it being a late-time modification), breaking degeneracies with $\sigma_8$ and $\sum m_\nu$. This means more redshift information will improve constraints, particularly high redshift information which was also pointed out in \cite{Harnois-Deraps:2022bie}, where they showed that most Fisher information on modified gravity is contained in the highest redshift bin. Finally, it was shown in \cite{Taylor:2018nrc} that a large number of bins are needed to capture the high-redshift information when using equipopulated binning. 

Lastly, a recent analysis \citep{Arico:2023ocu} of the Dark Energy Survey year 3 data including baryons, massive neutrinos and the same intrinsic alignment model employed here has shown consistent constraints on the $S_8$ parameter after considering they employ a sky fraction that is roughly 4 times less than that employed here. Their constraints on $\Omega_m$ are still significantly worse than ours, and the forecast of \cite{Euclid:2019clj} for example. We also employ 6 times the number density of their analysis, which will vastly improve constraining power on small scales. Survey specifications and the full power of Stage IV make such comparisons difficult, but we deem these results to be well within our expectations. It is also worth noting that this analysis \citep{Arico:2023ocu} was possible due to improvements in the prior ranges of the \href{https://baccoemu.readthedocs.io/en/latest/}{BACCO} emulator. Priors can be informative and it is important to have accurate nonlinear modelling across a wide parameter space \citep[see][for example]{Carrilho:2022mon}, making \react{} emulation an easy and viable means of performing data analyses for beyond-LCDM scenarios.

We conclude by noting that such a high dimensional parameter space analysis is both necessary and challenging when looking to optimise reliable information extraction from Stage IV surveys. It is also worthwhile as nonlinear scales can provide considerable information on interesting extensions to LCDM. Our analyses has focused on a minimal extension to LCDM, and serves as a `proof of principle' and `proof of worth' for the tools used in this paper, notably \cosmopower{} and \react{}, which we combined to provide \reactemufr{}. One key goal of forthcoming Stage IV surveys is to provide clues for solving the looming theoretical issues of Dark Energy and, even more broadly, the cosmological constant problem \cite[see][for a recent review on modified gravity solutions]{Bernardo:2022cck}. Thus, pipelines able to accurately model comprehensive extensions to LCDM becomes an absolute necessity. Such extensions have already been proposed \cite[see][for example]{Bose:2022vwi}, but include a number of new degrees of freedom, making the need for emulation even more clear, as is the development of fast inference pipelines based on gradient-based sampling techniques \citep[see \textit{e.g.}][]{Ruiz_Zapatero_2023, Campagne_2023, Piras2023}. It is straightforward to include existing models of gravity and dark energy into this emulation framework. It is also the goal of ongoing work to integrate these and perform similar tests as part of necessary preparation for the data releases of the largest galaxy surveys to date.

%
% \begin{figure*}
%     \centering
%     \includegraphics[width=\textwidth, height=0.95\textheight]{plots/lcdm_cosmo.png}
%     \caption{2-dimensional cosmological parameter posteriors for the $7\times2$ ({\bf bottom-left}) and $3\times2$ ({\bf top-right}) baryonic parameter models in {\bf Scenario A} (LCDM data vector). We show the $\ell_{\rm max}=1500,3000,5000$ cases for both baryonic models.}
%     \label{fig:cosmo_params_lcdm}
% \end{figure*}
%
\vspace{1cm}

% ================================================
%                   Acknowledgements
% ================================================

\section*{Acknowledgements}
\noindent The authors would like to thank A. Pourtsidou and L. Lombriser for useful comments. ASM is supported by the MSSL STFC Consolidated Grant ST/W001136/1. BB is supported by a UKRI Stephen Hawking Fellowship (EP/W005654/2) and acknowledges support from the Swiss National Science Foundation (SNSF) Professorship grant Nos.~170547 \& 202671. We acknowledge use of \texttt{GetDist} \citep{Lewis:2019xzd} for plotting contours. This work used facilities provided by the UCL Cosmoparticle Initiative. For the purpose of open access, the authors have applied a Creative Commons Attribution (CC BY) licence to any Author Accepted Manuscript version arising from this submission.

%%%%%%%%%%%%%%%%%%%%%%%%%%%%%%%%%%%%%%%%%%%%%%%%%%
\section*{Data Availability}
\reactemufr{} 
 is publicly available \href{https://github.com/cosmopower-organization/reactemu-fr}{\faicon{github}}.



\bibliographystyle{mnras}
\bibliography{main} % 

\begin{thebibliography}{}
\makeatletter
\relax
\def\mn@urlcharsother{\let\do\@makeother \do\$\do\&\do\#\do\^\do\_\do\%\do\~}
\def\mn@doi{\begingroup\mn@urlcharsother \@ifnextchar [ {\mn@doi@}
  {\mn@doi@[]}}
\def\mn@doi@[#1]#2{\def\@tempa{#1}\ifx\@tempa\@empty \href
  {http://dx.doi.org/#2} {doi:#2}\else \href {http://dx.doi.org/#2} {#1}\fi
  \endgroup}
\def\mn@eprint#1#2{\mn@eprint@#1:#2::\@nil}
\def\mn@eprint@arXiv#1{\href {http://arxiv.org/abs/#1} {{\tt arXiv:#1}}}
\def\mn@eprint@dblp#1{\href {http://dblp.uni-trier.de/rec/bibtex/#1.xml}
  {dblp:#1}}
\def\mn@eprint@#1:#2:#3:#4\@nil{\def\@tempa {#1}\def\@tempb {#2}\def\@tempc
  {#3}\ifx \@tempc \@empty \let \@tempc \@tempb \let \@tempb \@tempa \fi \ifx
  \@tempb \@empty \def\@tempb {arXiv}\fi \@ifundefined
  {mn@eprint@\@tempb}{\@tempb:\@tempc}{\expandafter \expandafter \csname
  mn@eprint@\@tempb\endcsname \expandafter{\@tempc}}}

\bibitem[\protect\citeauthoryear{Adamek et~al.}{Adamek
  et~al.}{2022}]{Euclid:2022qde}
Adamek J.,  et~al., 2022, {Euclid: Modelling massive neutrinos in cosmology --
  a code comparison} (\mn@eprint {arXiv} {2211.12457})

\bibitem[\protect\citeauthoryear{Aghanim et~al.}{Aghanim
  et~al.}{2020}]{Aghanim:2018eyx}
Aghanim N.,  et~al., 2020, \mn@doi [Astron. Astrophys.]
  {10.1051/0004-6361/201833910}, 641, A6

\bibitem[\protect\citeauthoryear{Ahmed et~al.}{Ahmed
  et~al.}{2004}]{SNO:2003bmh}
Ahmed S.~N.,  et~al., 2004, \mn@doi [Phys. Rev. Lett.]
  {10.1103/PhysRevLett.92.181301}, 92, 181301

\bibitem[\protect\citeauthoryear{Aker et~al.}{Aker
  et~al.}{2022}]{KATRIN:2021uub}
Aker M.,  et~al., 2022, \mn@doi [Nature Phys.] {10.1038/s41567-021-01463-1},
  18, 160

\bibitem[\protect\citeauthoryear{Amendola et~al.}{Amendola
  et~al.}{2018}]{Amendola:2016saw}
Amendola L.,  et~al., 2018, \mn@doi [Living Rev. Rel.]
  {10.1007/s41114-017-0010-3}, 21, 2

\bibitem[\protect\citeauthoryear{Angulo, Zennaro, Contreras, Aric\`o,
  Pellejero-Iba\~nez  \& St\"ucker}{Angulo et~al.}{2021}]{Angulo:2020vky}
Angulo R.~E.,  Zennaro M.,  Contreras S.,  Aric\`o G.,  Pellejero-Iba\~nez M.,
   St\"ucker J.,  2021, \mn@doi [Mon. Not. Roy. Astron. Soc.]
  {10.1093/mnras/stab2018}, 507, 5869

\bibitem[\protect\citeauthoryear{Aric\`o, Angulo, Contreras, Ondaro-Mallea,
  Pellejero-Iba\~nez  \& Zennaro}{Aric\`o et~al.}{2021}]{Arico:2020lhq}
Aric\`o G.,  Angulo R.~E.,  Contreras S.,  Ondaro-Mallea L.,
  Pellejero-Iba\~nez M.,   Zennaro M.,  2021, \mn@doi [Mon. Not. Roy. Astron.
  Soc.] {10.1093/mnras/stab1911}, 506, 4070

\bibitem[\protect\citeauthoryear{Aric\`o, Angulo, Zennaro, Contreras, Chen  \&
  Hern\'andez-Monteagudo}{Aric\`o et~al.}{2023}]{Arico:2023ocu}
Aric\`o G.,  Angulo R.~E.,  Zennaro M.,  Contreras S.,  Chen A.,
  Hern\'andez-Monteagudo C.,  2023, {DES Y3 cosmic shear down to small scales:
  constraints on cosmology and baryons} (\mn@eprint {arXiv} {2303.05537})

\bibitem[\protect\citeauthoryear{Arnold, Li, Giblin, Harnois-D\'eraps  \&
  Cai}{Arnold et~al.}{2021}]{Arnold:2021xtm}
Arnold C.,  Li B.,  Giblin B.,  Harnois-D\'eraps J.,   Cai Y.-C.,  2021, {}
  (\mn@eprint {arXiv} {2109.04984})

\bibitem[\protect\citeauthoryear{Asgari et~al.,}{Asgari
  et~al.}{2021}]{Asgari_2021}
Asgari M.,  et~al., 2021, \mn@doi [Astronomy {\&} Astrophysics]
  {10.1051/0004-6361/202039070}, 645, A104

\bibitem[\protect\citeauthoryear{Asgari, Mead  \& Heymans}{Asgari
  et~al.}{2023}]{Asgari_23}
Asgari M.,  Mead A.~J.,   Heymans C.,  2023, The halo model for cosmology: a
  pedagogical review (\mn@eprint {arXiv} {2303.08752})

\bibitem[\protect\citeauthoryear{{Bartelmann} \& {Schneider}}{{Bartelmann} \&
  {Schneider}}{2001}]{Bartelmann_2001}
{Bartelmann} M.,  {Schneider} P.,  2001, \mn@doi [\physrep]
  {10.1016/S0370-1573(00)00082-X}, \href
  {https://ui.adsabs.harvard.edu/abs/2001PhR...340..291B} {340, 291}

\bibitem[\protect\citeauthoryear{Bernardeau, Nishimichi  \& Taruya}{Bernardeau
  et~al.}{2014}]{Bernardeau_2014}
Bernardeau F.,  Nishimichi T.,   Taruya A.,  2014, \mn@doi [Monthly Notices of
  the Royal Astronomical Society] {10.1093/mnras/stu1861}, 445, 1526

\bibitem[\protect\citeauthoryear{Bernardo, Bose, Franzmann, Hagstotz, He, Litsa
   \& Niedermann}{Bernardo et~al.}{2023}]{Bernardo:2022cck}
Bernardo H.,  Bose B.,  Franzmann G.,  Hagstotz S.,  He Y.,  Litsa A.,
  Niedermann F.,  2023, \mn@doi [Universe] {10.3390/universe9020063}, 9, 63

\bibitem[\protect\citeauthoryear{Blanchard et~al.}{Blanchard
  et~al.}{2020}]{Euclid:2019clj}
Blanchard A.,  et~al., 2020, \mn@doi [Astron. Astrophys.]
  {10.1051/0004-6361/202038071}, 642, A191

\bibitem[\protect\citeauthoryear{Bose, Cataneo, Tr\"oster, Xia, Heymans  \&
  Lombriser}{Bose et~al.}{2020}]{Bose:2020wch}
Bose B.,  Cataneo M.,  Tr\"oster T.,  Xia Q.,  Heymans C.,   Lombriser L.,
  2020, \mn@doi [Mon. Not. Roy. Astron. Soc.] {10.1093/mnras/staa2696}, 498,
  4650

\bibitem[\protect\citeauthoryear{Bose et~al.,}{Bose
  et~al.}{2021}]{Bose:2021mkz}
Bose B.,  et~al., 2021, \mn@doi [Mon. Not. Roy. Astron. Soc.]
  {10.1093/mnras/stab2731}, 508, 2479

\bibitem[\protect\citeauthoryear{Bose, Tsedrik, Kennedy, Lombriser, Pourtsidou
  \& Taylor}{Bose et~al.}{2022}]{Bose:2022vwi}
Bose B.,  Tsedrik M.,  Kennedy J.,  Lombriser L.,  Pourtsidou A.,   Taylor A.,
  2022, {} (\mn@eprint {arXiv} {2210.01094})

\bibitem[\protect\citeauthoryear{Brax, Casas, Desmond  \& Elder}{Brax
  et~al.}{2021}]{Brax:2021wcv}
Brax P.,  Casas S.,  Desmond H.,   Elder B.,  2021, \mn@doi [Universe]
  {10.3390/universe8010011}, 8, 11

\bibitem[\protect\citeauthoryear{Bridle \& King}{Bridle \&
  King}{2007}]{Bridle:2007ft}
Bridle S.,  King L.,  2007, \mn@doi [New J. Phys.]
  {10.1088/1367-2630/9/12/444}, 9, 444

\bibitem[\protect\citeauthoryear{Campagne et~al.,}{Campagne
  et~al.}{2023}]{Campagne_2023}
Campagne J.-E.,  et~al., 2023, \mn@doi [The Open Journal of Astrophysics]
  {10.21105/astro.2302.05163}, 6

\bibitem[\protect\citeauthoryear{Carrilho, Carrion, Bose, Pourtsidou, Hidalgo,
  Lombriser  \& Baldi}{Carrilho et~al.}{2022}]{Carrilho:2021rqo}
Carrilho P.,  Carrion K.,  Bose B.,  Pourtsidou A.,  Hidalgo J.~C.,  Lombriser
  L.,   Baldi M.,  2022, \mn@doi [Mon. Not. Roy. Astron. Soc.]
  {10.1093/mnras/stac641}, 512, 3691

\bibitem[\protect\citeauthoryear{Carrilho, Moretti  \& Pourtsidou}{Carrilho
  et~al.}{2023}]{Carrilho:2022mon}
Carrilho P.,  Moretti C.,   Pourtsidou A.,  2023, \mn@doi [JCAP]
  {10.1088/1475-7516/2023/01/028}, 01, 028

\bibitem[\protect\citeauthoryear{Cataneo et~al.,}{Cataneo
  et~al.}{2015}]{Cataneo:2014kaa}
Cataneo M.,  et~al., 2015, \mn@doi [Phys.\ Rev.\ D]
  {10.1103/PhysRevD.92.044009}, 92, 044009

\bibitem[\protect\citeauthoryear{Cataneo, Lombriser, Heymans, Mead, Barreira,
  Bose  \& Li}{Cataneo et~al.}{2019}]{Cataneo:2018cic}
Cataneo M.,  Lombriser L.,  Heymans C.,  Mead A.,  Barreira A.,  Bose S.,   Li
  B.,  2019, \mn@doi [Mon.\ Not.\ Roy.\ Astron.\ Soc.] {10.1093/mnras/stz1836},
  488, 2121

\bibitem[\protect\citeauthoryear{Copeland, Taylor  \& Hall}{Copeland
  et~al.}{2018}]{Copeland:2017hzu}
Copeland D.,  Taylor A.,   Hall A.,  2018, \mn@doi [Mon. Not. Roy. Astron.
  Soc.] {10.1093/mnras/sty2001}, 480, 2247

\bibitem[\protect\citeauthoryear{Deshpande, Taylor  \& Kitching}{Deshpande
  et~al.}{2020}]{Deshpande_20}
Deshpande A.~C.,  Taylor P.~L.,   Kitching T.~D.,  2020, \mn@doi [Phys. Rev. D]
  {10.1103/PhysRevD.102.083535}, 102, 083535

\bibitem[\protect\citeauthoryear{Desmond \& Ferreira}{Desmond \&
  Ferreira}{2020}]{Desmond:2020gzn}
Desmond H.,  Ferreira P.~G.,  2020, \mn@doi [Phys. Rev. D]
  {10.1103/PhysRevD.102.104060}, 102, 104060

\bibitem[\protect\citeauthoryear{Fukuda et~al.}{Fukuda
  et~al.}{1998}]{Super-Kamiokande:1998qwk}
Fukuda Y.,  et~al., 1998, \mn@doi [Phys. Rev. Lett.]
  {10.1103/PhysRevLett.81.1158}, 81, 1158

\bibitem[\protect\citeauthoryear{Giblin, Cataneo, Moews  \& Heymans}{Giblin
  et~al.}{2019}]{Giblin:2019iit}
Giblin B.,  Cataneo M.,  Moews B.,   Heymans C.,  2019, \mn@doi [Mon. Not. Roy.
  Astron. Soc.] {10.1093/mnras/stz2659}, 490, 4826

\bibitem[\protect\citeauthoryear{Giri \& Schneider}{Giri \&
  Schneider}{2021}]{Giri:2021qin}
Giri S.~K.,  Schneider A.,  2021, \mn@doi [JCAP]
  {10.1088/1475-7516/2021/12/046}, 12, 046

\bibitem[\protect\citeauthoryear{Handley, Hobson  \& Lasenby}{Handley
  et~al.}{2015a}]{Handley_2015a}
Handley W.~J.,  Hobson M.~P.,   Lasenby A.~N.,  2015a, \mn@doi [Monthly Notices
  of the Royal Astronomical Society: Letters] {10.1093/mnrasl/slv047}, 450, L61

\bibitem[\protect\citeauthoryear{Handley, Hobson  \& Lasenby}{Handley
  et~al.}{2015b}]{Handley_2015b}
Handley W.~J.,  Hobson M.~P.,   Lasenby A.~N.,  2015b, \mn@doi [Monthly Notices
  of the Royal Astronomical Society] {10.1093/mnras/stv1911}, 453, 4385

\bibitem[\protect\citeauthoryear{Harnois-D\'eraps, Hernandez-Aguayo,
  Cuesta-Lazaro, Arnold, Li, Davies  \& Cai}{Harnois-D\'eraps
  et~al.}{2022}]{Harnois-Deraps:2022bie}
Harnois-D\'eraps J.,  Hernandez-Aguayo C.,  Cuesta-Lazaro C.,  Arnold C.,  Li
  B.,  Davies C.~T.,   Cai Y.-C.,  2022, {} (\mn@eprint {arXiv} {2211.05779})

\bibitem[\protect\citeauthoryear{Hu \& Sawicki}{Hu \&
  Sawicki}{2007}]{Hu:2007nk}
Hu W.,  Sawicki I.,  2007, \mn@doi [Phys.Rev.] {10.1103/PhysRevD.76.064004},
  D76, 064004

\bibitem[\protect\citeauthoryear{Huang et~al.}{Huang
  et~al.}{2021}]{DES:2020rmk}
Huang H.-J.,  et~al., 2021, \mn@doi [Mon. Not. Roy. Astron. Soc.]
  {10.1093/mnras/stab357}, 502, 6010

\bibitem[\protect\citeauthoryear{Ivezi\'c et~al.}{Ivezi\'c
  et~al.}{2019}]{LSST:2008ijt}
Ivezi\'c v.,  et~al., 2019, \mn@doi [Astrophys. J.] {10.3847/1538-4357/ab042c},
  873, 111

\bibitem[\protect\citeauthoryear{{Joachimi} \& {Bridle}}{{Joachimi} \&
  {Bridle}}{2010}]{Joachimi_2010}
{Joachimi} B.,  {Bridle} S.~L.,  2010, \mn@doi [\aap]
  {10.1051/0004-6361/200913657}, \href
  {https://ui.adsabs.harvard.edu/abs/2010A&A...523A...1J} {523, A1}

\bibitem[\protect\citeauthoryear{Joachimi, Schneider  \& Eifler}{Joachimi
  et~al.}{2007}]{Joachimi_2007}
Joachimi B.,  Schneider P.,   Eifler T.,  2007, \mn@doi [Astronomy {\&}
  Astrophysics] {10.1051/0004-6361:20078400}, 477, 43

\bibitem[\protect\citeauthoryear{Kaiser}{Kaiser}{1998}]{Kaiser_1998}
Kaiser N.,  1998, \mn@doi [The Astrophysical Journal] {10.1086/305515}, 498, 26

\bibitem[\protect\citeauthoryear{{Kilbinger}}{{Kilbinger}}{2015}]{Kilbinger_2015}
{Kilbinger} M.,  2015, \mn@doi [Reports on Progress in Physics]
  {10.1088/0034-4885/78/8/086901}, \href
  {https://ui.adsabs.harvard.edu/abs/2015RPPh...78h6901K} {78, 086901}

\bibitem[\protect\citeauthoryear{Kilbinger et~al.,}{Kilbinger
  et~al.}{2017}]{Kilbinger_2017}
Kilbinger M.,  et~al., 2017, \mn@doi [Monthly Notices of the Royal Astronomical
  Society] {10.1093/mnras/stx2082}, 472, 2126

\bibitem[\protect\citeauthoryear{Kitching, Alsing, Heavens, Jimenez, McEwen  \&
  Verde}{Kitching et~al.}{2017}]{Kitching_2017}
Kitching T.~D.,  Alsing J.,  Heavens A.~F.,  Jimenez R.,  McEwen J.~D.,   Verde
  L.,  2017, \mn@doi [Monthly Notices of the Royal Astronomical Society]
  {10.1093/mnras/stx1039}, 469, 2737

\bibitem[\protect\citeauthoryear{Knabenhans et~al.}{Knabenhans
  et~al.}{2019}]{Euclid:2018mlb}
Knabenhans M.,  et~al., 2019, \mn@doi [Mon. Not. Roy. Astron. Soc.]
  {10.1093/mnras/stz197}, 484, 5509

\bibitem[\protect\citeauthoryear{Knabenhans et~al.}{Knabenhans
  et~al.}{2021}]{Euclid:2020rfv}
Knabenhans M.,  et~al., 2021, \mn@doi [Mon. Not. Roy. Astron. Soc.]
  {10.1093/mnras/stab1366}, 505, 2840

\bibitem[\protect\citeauthoryear{Koyama}{Koyama}{2016}]{Koyama_2016}
Koyama K.,  2016, \mn@doi [Reports on Progress in Physics]
  {10.1088/0034-4885/79/4/046902}, 79, 046902

\bibitem[\protect\citeauthoryear{Krause \& Eifler}{Krause \&
  Eifler}{2017}]{Krause_2017}
Krause E.,  Eifler T.,  2017, \mn@doi [Monthly Notices of the Royal
  Astronomical Society] {10.1093/mnras/stx1261}, 470, 2100

\bibitem[\protect\citeauthoryear{{Lacasa}, {Lima}  \& {Aguena}}{{Lacasa}
  et~al.}{2018}]{Lacasa_2018}
{Lacasa} F.,  {Lima} M.,   {Aguena} M.,  2018, \mn@doi [\aap]
  {10.1051/0004-6361/201630281}, \href
  {https://ui.adsabs.harvard.edu/abs/2018A&A...611A..83L} {611, A83}

\bibitem[\protect\citeauthoryear{Laureijs et~al.}{Laureijs
  et~al.}{2011}]{EUCLID:2011zbd}
Laureijs R.,  et~al., 2011, {} (\mn@eprint {arXiv} {1110.3193})

\bibitem[\protect\citeauthoryear{Lawrence et~al.,}{Lawrence
  et~al.}{2017}]{Lawrence:2017ost}
Lawrence E.,  et~al., 2017, \mn@doi [Astrophys. J.] {10.3847/1538-4357/aa86a9},
  847, 50

\bibitem[\protect\citeauthoryear{Lesgourgues \& Pastor}{Lesgourgues \&
  Pastor}{2006}]{Lesgourgues:2006nd}
Lesgourgues J.,  Pastor S.,  2006, \mn@doi [Phys. Rept.]
  {10.1016/j.physrep.2006.04.001}, 429, 307

\bibitem[\protect\citeauthoryear{Lewis}{Lewis}{2019}]{Lewis:2019xzd}
Lewis A.,  2019, {} (\mn@eprint {arXiv} {1910.13970})

\bibitem[\protect\citeauthoryear{Li, Hu  \& Takada}{Li et~al.}{2014}]{Li_2014}
Li Y.,  Hu W.,   Takada M.,  2014, \mn@doi [Physical Review D]
  {10.1103/physrevd.89.083519}, 89

\bibitem[\protect\citeauthoryear{LoVerde \& Afshordi}{LoVerde \&
  Afshordi}{2008}]{LoVerde_2008}
LoVerde M.,  Afshordi N.,  2008, \mn@doi [Phys. Rev. D]
  {10.1103/PhysRevD.78.123506}, 78, 123506

\bibitem[\protect\citeauthoryear{Lombriser}{Lombriser}{2014}]{Lombriser:2014dua}
Lombriser L.,  2014, \mn@doi [Annalen Phys.] {10.1002/andp.201400058}, 526, 259

\bibitem[\protect\citeauthoryear{Martinelli et~al.}{Martinelli
  et~al.}{2021}]{Euclid:2020tff}
Martinelli M.,  et~al., 2021, \mn@doi [Astron. Astrophys.]
  {10.1051/0004-6361/202039835}, 649, A100

\bibitem[\protect\citeauthoryear{Mead}{Mead}{2017}]{Mead:2016ybv}
Mead A.,  2017, \mn@doi [Mon. Not. Roy. Astron. Soc.] {10.1093/mnras/stw2312},
  464, 1282

\bibitem[\protect\citeauthoryear{Mead, Heymans, Lombriser, Peacock, Steele  \&
  Winther}{Mead et~al.}{2016}]{Mead:2016zqy}
Mead A.,  Heymans C.,  Lombriser L.,  Peacock J.,  Steele O.,   Winther H.,
  2016, \mn@doi [Mon. Not. Roy. Astron. Soc.] {10.1093/mnras/stw681}, 459, 1468

\bibitem[\protect\citeauthoryear{Mead, Brieden, Tr\"oster  \& Heymans}{Mead
  et~al.}{2020}]{Mead:2020vgs}
Mead A.,  Brieden S.,  Tr\"oster T.,   Heymans C.,  2020, \mn@doi [Mon. Not.
  Roy. Astron. Soc.] {10.1093/mnras/stab082}

\bibitem[\protect\citeauthoryear{Merloni et~al.}{Merloni
  et~al.}{2012}]{eROSITA:2012lfj}
Merloni A.,  et~al., 2012, {eROSITA Science Book: Mapping the Structure of the
  Energetic Universe} (\mn@eprint {arXiv} {1209.3114})

\bibitem[\protect\citeauthoryear{Parimbelli, Carbone, Bel, Bose, Calabrese,
  Carella  \& Zennaro}{Parimbelli et~al.}{2022}]{Parimbelli:2022pmr}
Parimbelli G.,  Carbone C.,  Bel J.,  Bose B.,  Calabrese M.,  Carella E.,
  Zennaro M.,  2022, \mn@doi [JCAP] {10.1088/1475-7516/2022/11/041}, 11, 041

\bibitem[\protect\citeauthoryear{{Piras} \& {Spurio Mancini}}{{Piras} \&
  {Spurio Mancini}}{2023}]{Piras2023}
{Piras} D.,  {Spurio Mancini} A.,  2023, \mn@doi [The Open Journal of
  Astrophysics] {10.21105/astro.2305.06347}, 6

\bibitem[\protect\citeauthoryear{Ramachandra, Valogiannis, Ishak  \&
  Heitmann}{Ramachandra et~al.}{2021}]{Ramachandra:2020lue}
Ramachandra N.,  Valogiannis G.,  Ishak M.,   Heitmann K.,  2021, \mn@doi
  [Phys. Rev. D] {10.1103/PhysRevD.103.123525}, 103, 123525

\bibitem[\protect\citeauthoryear{Ruiz-Zapatero, Hadzhiyska, Alonso, Ferreira,
  Garc{\'{\i}}a-Garc{\'{\i}}a  \& Mootoovaloo}{Ruiz-Zapatero
  et~al.}{2023}]{Ruiz_Zapatero_2023}
Ruiz-Zapatero J.,  Hadzhiyska B.,  Alonso D.,  Ferreira P.~G.,
  Garc{\'{\i}}a-Garc{\'{\i}}a C.,   Mootoovaloo A.,  2023, \mn@doi [Monthly
  Notices of the Royal Astronomical Society] {10.1093/mnras/stad1192}

\bibitem[\protect\citeauthoryear{Samuroff, Mandelbaum  \& Blazek}{Samuroff
  et~al.}{2021}]{Samuroff:2020gpm}
Samuroff S.,  Mandelbaum R.,   Blazek J.,  2021, \mn@doi [Mon. Not. Roy.
  Astron. Soc.] {10.1093/mnras/stab2520}, 508, 637

\bibitem[\protect\citeauthoryear{Sato \& Nishimichi}{Sato \&
  Nishimichi}{2013}]{Sato_2013}
Sato M.,  Nishimichi T.,  2013, \mn@doi [Physical Review D]
  {10.1103/physrevd.87.123538}, 87

\bibitem[\protect\citeauthoryear{Schneider, Stoira, Refregier, Weiss,
  Knabenhans, Stadel  \& Teyssier}{Schneider et~al.}{2020a}]{Schneider:2019snl}
Schneider A.,  Stoira N.,  Refregier A.,  Weiss A.~J.,  Knabenhans M.,  Stadel
  J.,   Teyssier R.,  2020a, \mn@doi [JCAP] {10.1088/1475-7516/2020/04/019},
  04, 019

\bibitem[\protect\citeauthoryear{Schneider et~al.,}{Schneider
  et~al.}{2020b}]{Schneider:2019xpf}
Schneider A.,  et~al., 2020b, \mn@doi [JCAP] {10.1088/1475-7516/2020/04/020},
  04, 020

\bibitem[\protect\citeauthoryear{Semboloni, Hoekstra, Schaye, van Daalen  \&
  McCarthy}{Semboloni et~al.}{2011}]{Semboloni:2011fe}
Semboloni E.,  Hoekstra H.,  Schaye J.,  van Daalen M.~P.,   McCarthy I.~J.,
  2011, \mn@doi [Mon. Not. Roy. Astron. Soc.]
  {10.1111/j.1365-2966.2011.19385.x}, 417, 2020

\bibitem[\protect\citeauthoryear{Smail, Ellis, Fitchett  \& Edge}{Smail
  et~al.}{1995}]{Smail_1995}
Smail I.,  Ellis R.~S.,  Fitchett M.~J.,   Edge A.~C.,  1995, \mn@doi [Monthly
  Notices of the Royal Astronomical Society] {10.1093/mnras/273.2.277}, 273,
  277

\bibitem[\protect\citeauthoryear{{Spergel} et~al.,}{{Spergel}
  et~al.}{2015}]{Roman}
{Spergel} D.,  et~al., 2015, arXiv e-prints, \href
  {https://ui.adsabs.harvard.edu/abs/2015arXiv150303757S} {p. arXiv:1503.03757}

\bibitem[\protect\citeauthoryear{Spurio~Mancini \& Pourtsidou}{Spurio~Mancini
  \& Pourtsidou}{2022}]{ManciniSpurio:2021jvx}
Spurio~Mancini A.,  Pourtsidou A.,  2022, \mn@doi [Mon. Not. Roy. Astron. Soc.]
  {10.1093/mnrasl/slac019}, 512, L44

\bibitem[\protect\citeauthoryear{Spurio~Mancini, Piras, Alsing, Joachimi  \&
  Hobson}{Spurio~Mancini et~al.}{2022}]{SpurioMancini:2021ppk}
Spurio~Mancini A.,  Piras D.,  Alsing J.,  Joachimi B.,   Hobson M.~P.,  2022,
  \mn@doi [Mon. Not. Roy. Astron. Soc.] {10.1093/mnras/stac064}, 511, 1771

\bibitem[\protect\citeauthoryear{Srinivasan, Thomas, Pace  \&
  Battye}{Srinivasan et~al.}{2021}]{Srinivasan:2021gib}
Srinivasan S.,  Thomas D.~B.,  Pace F.,   Battye R.,  2021, \mn@doi [JCAP]
  {10.1088/1475-7516/2021/06/016}, 06, 016

\bibitem[\protect\citeauthoryear{Takada \& Hu}{Takada \&
  Hu}{2013}]{Takada_2013}
Takada M.,  Hu W.,  2013, \mn@doi [Physical Review D]
  {10.1103/physrevd.87.123504}, 87

\bibitem[\protect\citeauthoryear{Takahashi, Sato, Nishimichi, Taruya  \&
  Oguri}{Takahashi et~al.}{2012}]{Takahashi:2012em}
Takahashi R.,  Sato M.,  Nishimichi T.,  Taruya A.,   Oguri M.,  2012, \mn@doi
  [Astrophys. J.] {10.1088/0004-637X/761/2/152}, 761, 152

\bibitem[\protect\citeauthoryear{Taylor, Bernardeau  \& Kitching}{Taylor
  et~al.}{2018a}]{Taylor_18}
Taylor P.~L.,  Bernardeau F.,   Kitching T.~D.,  2018a, \mn@doi [Phys. Rev. D]
  {10.1103/PhysRevD.98.083514}, 98, 083514

\bibitem[\protect\citeauthoryear{Taylor, Kitching  \& McEwen}{Taylor
  et~al.}{2018b}]{Taylor:2018nrc}
Taylor P.~L.,  Kitching T.~D.,   McEwen J.~D.,  2018b, \mn@doi [Phys. Rev. D]
  {10.1103/PhysRevD.98.043532}, 98, 043532

\bibitem[\protect\citeauthoryear{Winther, Casas, Baldi, Koyama, Li, Lombriser
  \& Zhao}{Winther et~al.}{2019}]{Winther:2019mus}
Winther H.,  Casas S.,  Baldi M.,  Koyama K.,  Li B.,  Lombriser L.,   Zhao
  G.-B.,  2019, \mn@doi [Phys. Rev. D] {10.1103/PhysRevD.100.123540}, 100,
  123540

\bibitem[\protect\citeauthoryear{Wright, Koyama, Winther  \& Zhao}{Wright
  et~al.}{2019}]{Wright:2019qhf}
Wright B.~S.,  Koyama K.,  Winther H.~A.,   Zhao G.-B.,  2019, \mn@doi [JCAP]
  {10.1088/1475-7516/2019/06/040}, 06, 040

\bibitem[\protect\citeauthoryear{Zhao}{Zhao}{2014}]{Zhao:2013dza}
Zhao G.-B.,  2014, \mn@doi [Astrophys. J. Suppl.] {10.1088/0067-0049/211/2/23},
  211, 23

\makeatother
\end{thebibliography}

%%%%%%%%%%%%%%%%%%%%%%%%%%%%%%%%%%%%%%%%%%%%%%%%%%

%%%%%%%%%%%%%%%%% APPENDICES %%%%%%%%%%%%%%%%%%%%%

\appendix
 
\section{Full posteriors}
\label{app:fullpost}
In this appendix we present the full 2-dimensional marginalised posteriors for all analyses, with the exception of the baryonic feedback redshift-dependency parameters, which we have found to be largely prior dominated. 

\begin{figure*}
    \centering
    \includegraphics[width=\textwidth]{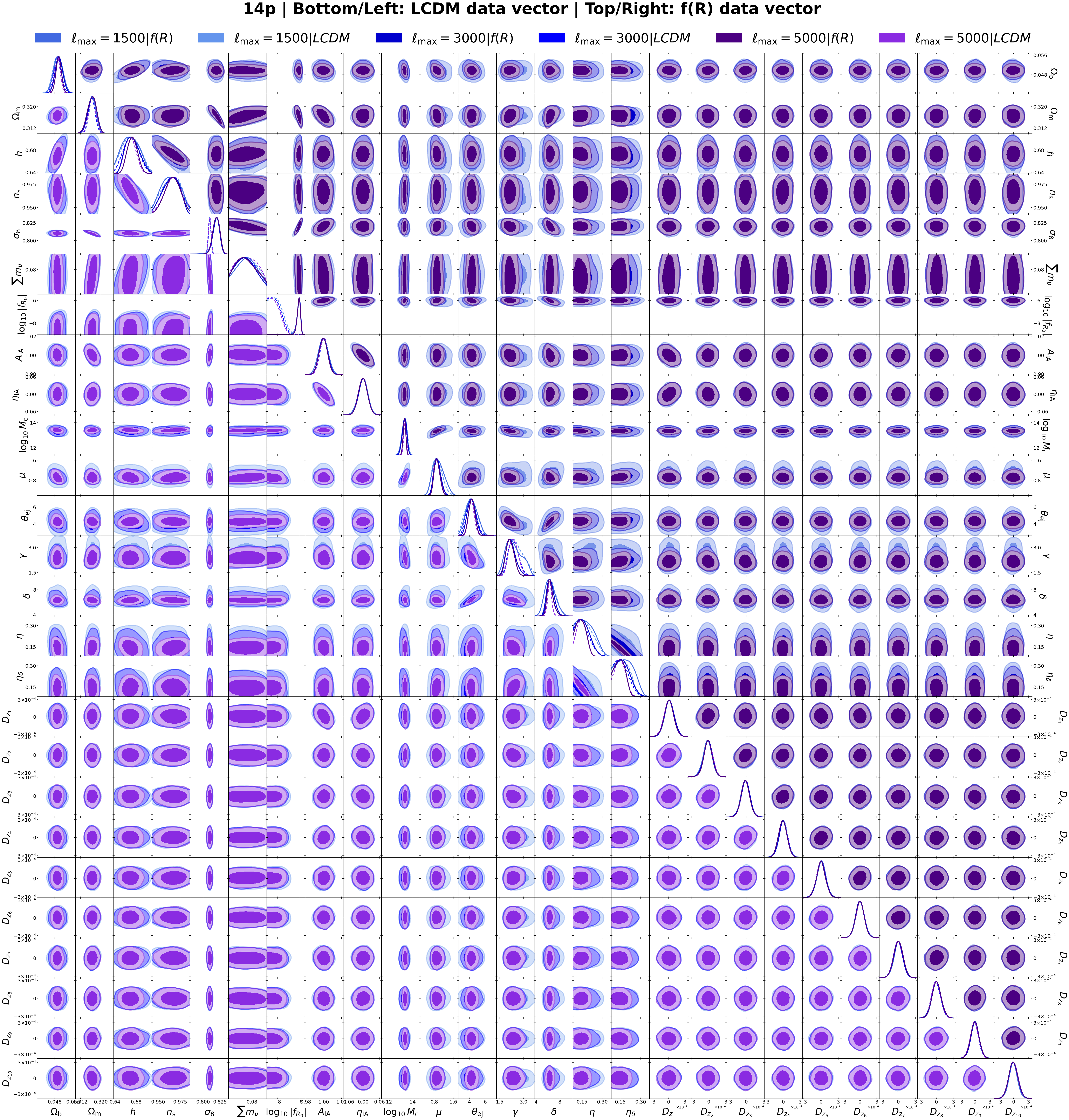}
    \caption{Full set of marginalised 2-dimensional posterior distributions for the $7\times2$ baryonic parameter model. We show all scale cuts as well as the case where we fit to a LCDM fiducial data vector ({\bf bottom left}) and $f(R)$ fiducial data vector ({\bf top right}).}
    \label{fig:all_14p}
\end{figure*}

\begin{figure*}
    \centering
    \includegraphics[width=\textwidth]{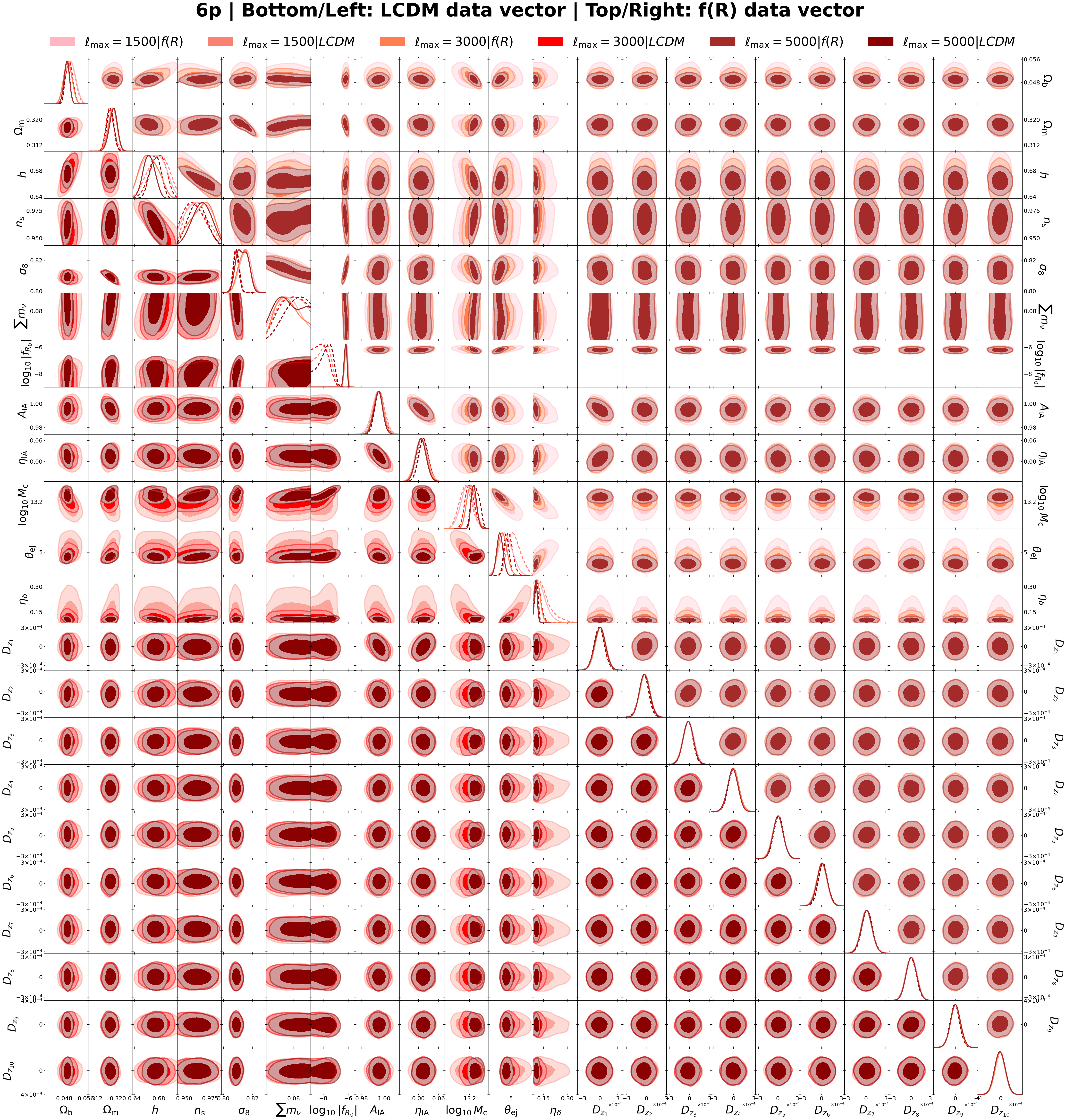}
    \caption{Same as \autoref{fig:all_14p} but for the $3\times2$ baryonic parameter model case.}
    \label{fig:all_6p}
\end{figure*}

In \autoref{fig:all_14p} we show the results for the $7\times2$ baryonic parameter model, both when considering a fiducial LCDM data vector and $f(R)$ data vector (Scenarios A \& B respectively in the main text). All scale cuts are also shown. Similarly, in \autoref{fig:all_6p} we show the results for the reduced $3\times2$ baryonic parameter model. 

These plots visually highlight the vast parameter space that one needs to probe in order to comprehensively account for all known physics, and probe minimal extensions to LCDM. This space can foreseeably increase when looking to move deeper into the nonlinear regime or when wishing to probe more general deviations to LCDM. 


\label{lastpage}
\end{document}